\documentclass{iopjournal}
\usepackage{natbib}
\usepackage{amsmath}
\usepackage{subfig}
\usepackage{tikz}
\usetikzlibrary{shapes.geometric, arrows, positioning, fit, matrix,arrows.meta, calc    }
\usepackage{booktabs} 
\usepackage{subcaption} 
\usepackage[normalem]{ulem}
\usepackage[pagewise]{lineno}
\usepackage{changes} 
\usepackage{subfig}

\setaddedmarkup{\textbf{#1}}
\setdeletedmarkup{}
\setauthormarkup{none}

\tikzset{
    inputlayer/.style={draw, fill=blue!20, rounded corners, minimum width=1.9cm, minimum height=0.7cm, align=center, font=\scriptsize},
    process/.style={draw, fill=green!20, rounded corners, minimum width=1.9cm, minimum height=0.7cm, align=center, font=\scriptsize},
    dense/.style={draw, fill=orange!20, rounded corners, minimum width=1.9cm, minimum height=0.7cm, align=center, font=\scriptsize},
    dropout/.style={draw, fill=red!20, rounded corners, minimum width=1.9cm, minimum height=0.7cm, align=center, font=\scriptsize},
    bigop/.style={draw, fill=purple!20, rounded corners, minimum width=4cm, minimum height=0.8cm, align=center, font=\scriptsize},
    arrow/.style={-{Latex[length=2mm,width=1.5mm]}, thick}
}

\begin{document}
\articletype{Paper} 

\title{Multivariate Time Series Classification of {\it {\it Fermi}}-Detected Gamma-Ray Transients Using Convolutional-Recurrent Neural Networks}

\author{Arpan Aryam John$^1$\orcid{0009-0004-0636-5032}, Krushna Govind Shete$^1$\orcid{0009-0000-7209-0517}, Shabnam Iyyani$^{1,2,*}$\orcid{0000-0002-2525-3464} and Saptarshi Bej$^{3}$\orcid{0000-0003-1835-6139}}

\affil{$^1$ School of Physics, Indian Institute of Science Education and Research Thiruvananthapuram, 695551, India}

\affil{$^2$ Centre of High Performance Computing, Indian Institute of Science Education and Research Thiruvananthapuram, 695551, India}

\affil{$^3$ School of Data Science, Indian Institute of Science Education and Research Thiruvananthapuram, 695551, India}

\affil{$^*$Author to whom any correspondence should be addressed.}

\email{shabnam@iisertvm.ac.in}

\keywords{High energy astrophysics, Gamma-ray astronomy, Deep Learning, Convolutional neural networks, Recurrent neural networks, LSTM}

\begin{abstract}
{\it Fermi} Gamma-ray Space Telescope has detected a diverse range of gamma-ray transients since its launch in 2008. Over the years, {\it Fermi} has accumulated an extensive public archive of transient events. Traditional classification methods for these events typically rely on fixed thresholds, localisation accuracy, and characteristic light-curve features. However, in the current era of time-critical, multi-wavelength, and multi-messenger astronomy, rapid and reliable classification is essential to enable timely follow-up and coordinated observations.

In this work, we develop and present two deep learning-based classifiers that integrate convolutional and recurrent neural network architectures. Using multivariate time-series inputs derived from {\it Fermi}-GBM data, our models are trained to distinguish among four classes of gamma-ray transients: Gamma-Ray Bursts (GRBs), Terrestrial Gamma-ray Flashes (TGFs), Solar Flares (SFLAREs), and Soft Gamma Repeaters (SGRs). Furthermore, the models are designed to flag events that do not conform to any of these categories, providing a pathway for identifying potentially new or rare transient types.

Training was conducted using a carefully curated subset of high-confidence {\it Fermi} events. The resulting models achieve an overall classification accuracy of $93\%$, and identify approximately $2.5\%$ of the triggers as outliers of unknown origin. When applied to {\it Fermi} events with uncertain classifications, our models assign $60\%$ of them to the TGF category with over $60\%$ confidence.

These results demonstrate that incorporating deep learning-based classification into onboard or automated data pipelines can significantly enhance transient identification, minimize misclassification, and improve the discovery potential of new phenomena in future high-energy astrophysics missions.
\end{abstract}

\section{Introduction}
\begin{figure*}[ht]
    \centering
    \begin{tabular}{cc}
        \includegraphics[width=0.45\textwidth]{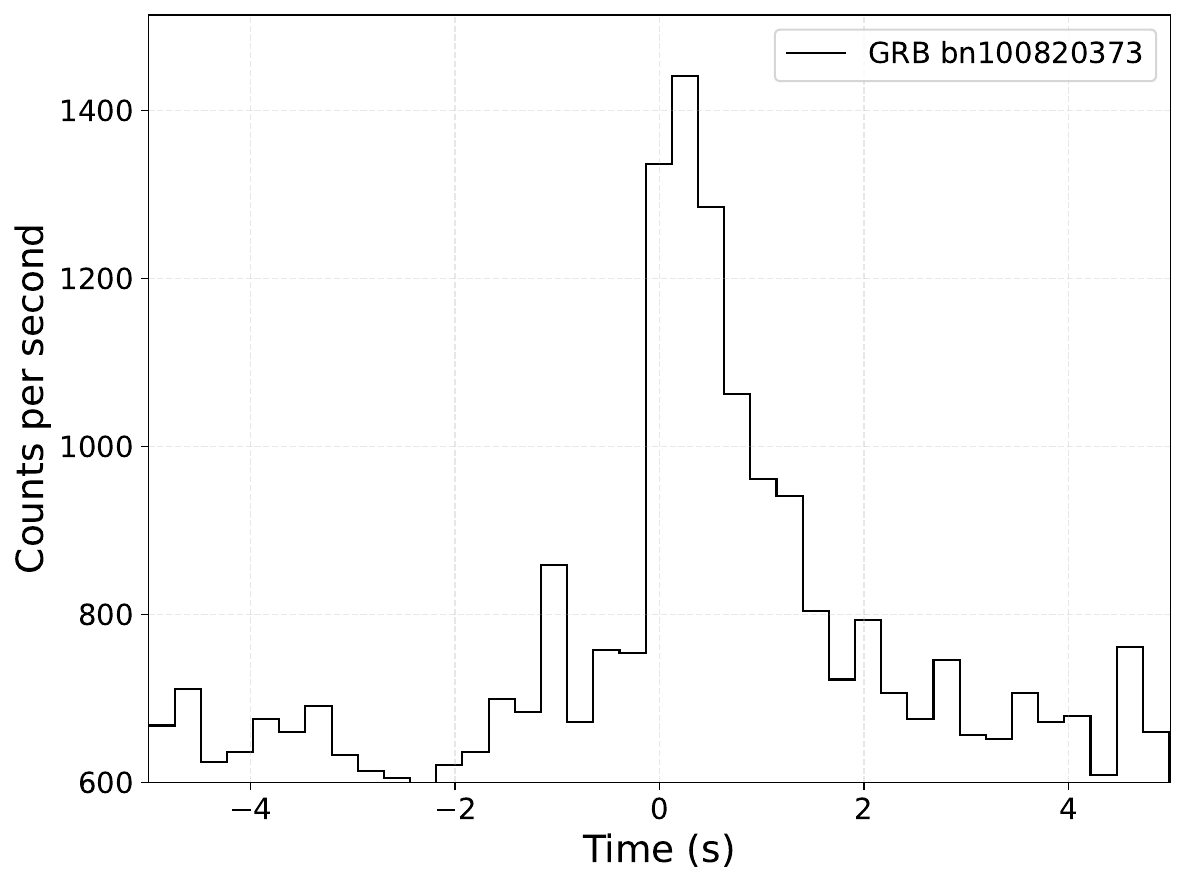} & 
        \includegraphics[width=0.45\textwidth]{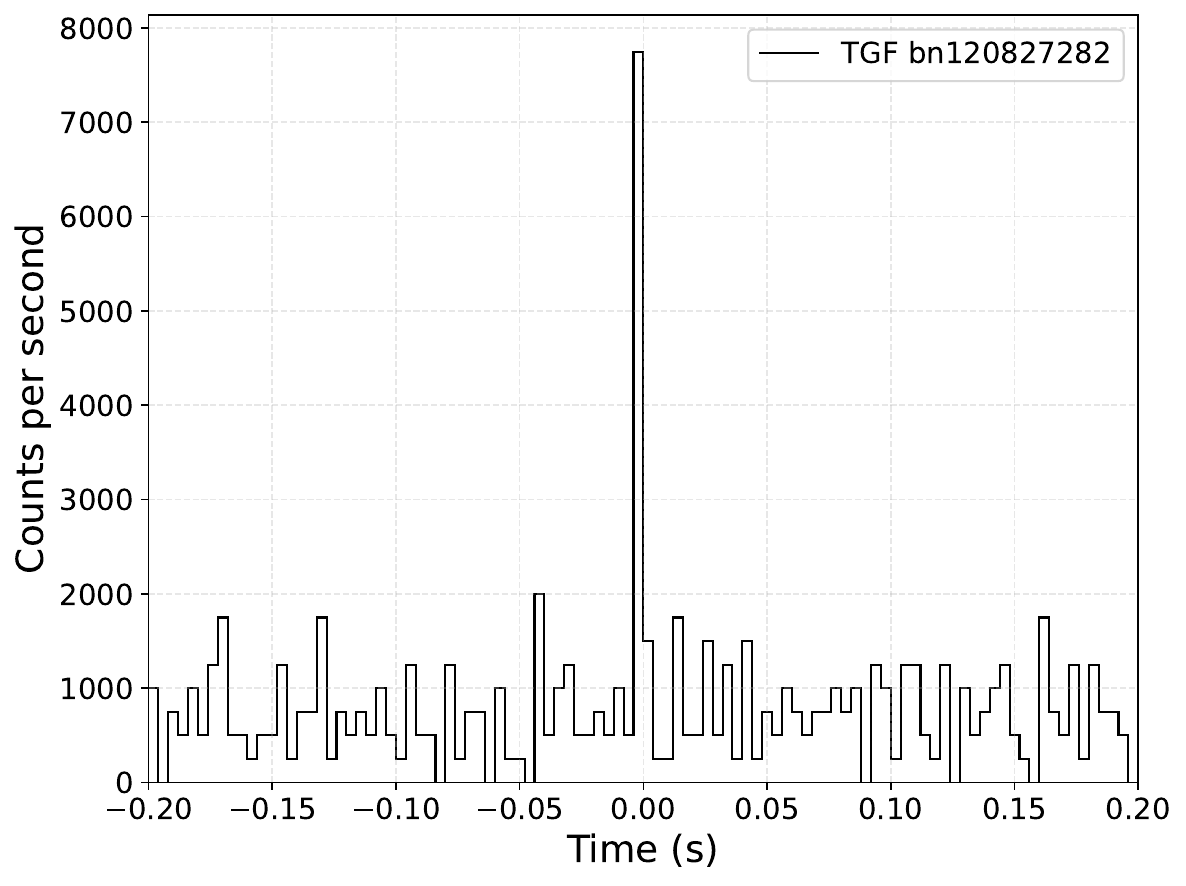} \\
         (a) GRB bn100820373 & (b) TGF bn120827282 
        \\
        \includegraphics[width=0.45\textwidth]{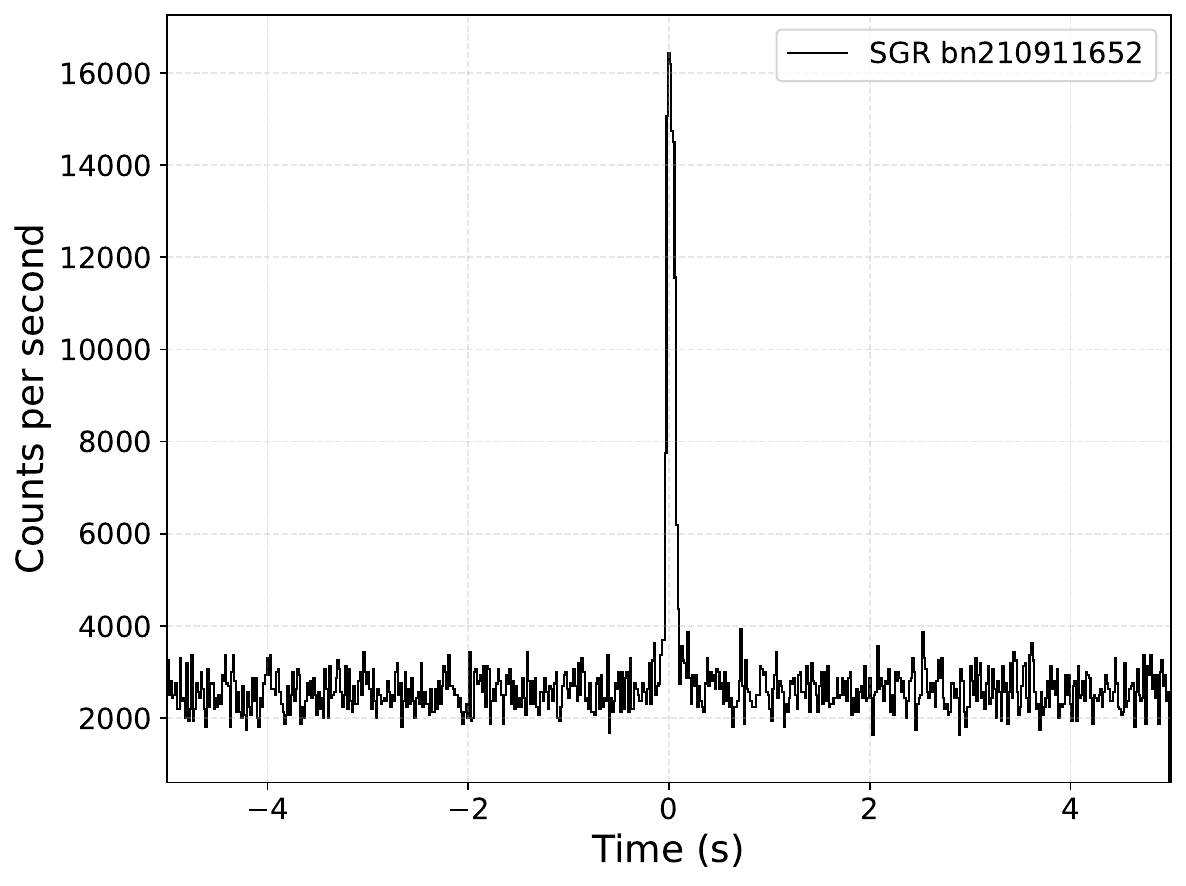} & 
        \includegraphics[width=0.45\textwidth]{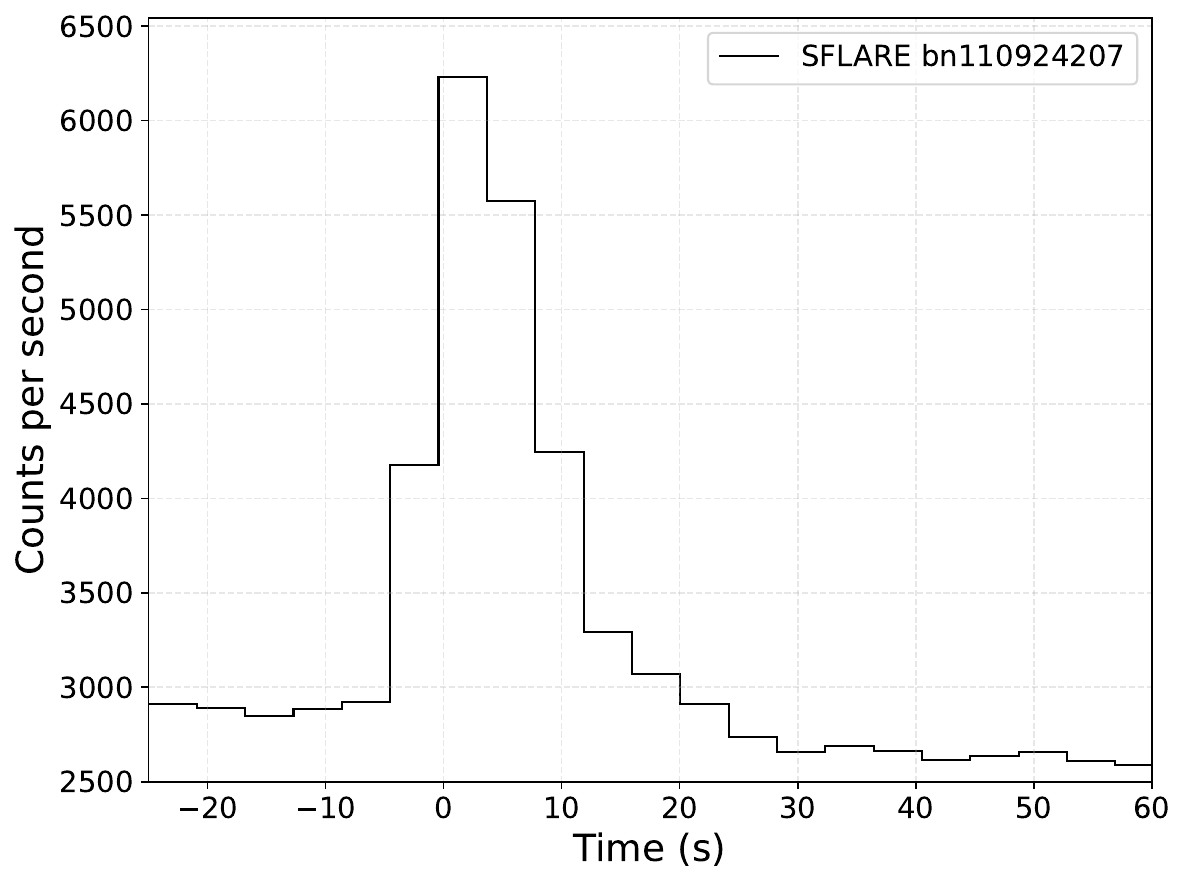} 
        \\
         (c) SGR bn210911652 & (d) SFLARE bn110924207
    \end{tabular}
    
    \caption{Examples of the lightcurves of different transient events with data taken from Fermi GBM, shown with different time bin sizes. (a) GRB bn100820373 with a bin size of 0.256 s in the 100\textendash900 keV energy range, displaying a characteristic burst profile with rapid variability. (b) TGF bn120827282 with a bin size of 0.004 s in the 100\textendash900 keV energy range, showing an extremely short-duration peak typical of terrestrial gamma\textendash ray flashes. (c) SGR bn210911652 with a bin size of 0.016 s in the 10\textendash100 keV energy range, exhibiting multiple short spikes, a signature of magnetar bursts. (d) SFLARE bn110924207 with a bin size of 4.096 s in the 10\textendash100 keV energy range, highlighting a gradual rise and decay structure often seen in solar flares.}
    \label{fig:lc_examples}
\end{figure*}
High energy gamma ray astrophysical transients are short-lived bursts of hard X-rays or gamma rays, 
that produce dramatic changes in intensity on time scales ranging from milliseconds to weeks. These are usually produced by the most extreme conditions in our universe, like massive 
stellar explosions, compact object mergers, magnetospheric reconnections in neutron stars, solar activities, etc, and, hence, are a tool for probing deep into the mechanisms behind our high energy Universe. Some of the extensively observed and studied gamma ray transients include Gamma Ray Bursts (GRBs, \citealt{Fermi_GRB_catalog_2020}), Terrestrial Gamma-ray Flashes (TGFs, \citealt{Fermi_TGF_catalog_2018,Agile_TGF_catalog_2020}), Solar Flares (SFLAREs, \citealt{Fermi_solar_flare_2014}), Soft Gamma ray Repeaters (SGRs, \citealt{SGR_Review_2011,SGR_catalog_Integral_2014}) etc. 

The Gamma-ray Burst Monitor (GBM) \citep{Meegan_2009} launched on June 11, 2008 onboard the {\it Fermi} Gamma-ray Space Telescope was designed to study high-energy transients, particularly gamma-ray bursts (GRBs).
It continuously monitors nearly the entire sky across a broad energy range from a few keV to several MeV. An extensive open source database known of detected and classified triggers is publicly available.


The automatic classification of transients detected by {\it Fermi} Gamma ray Burst Monitor (GBM) currently follows a two-step process. The first step involves an onboard classification performed almost instantly (1-5s) by the 
flight software\citep{Meegan_2009}. This system utilises a combination of event localisation, spectral hardness ratios, and McIlwain L-coordinates, 
applying Bayesian analysis to assign a probabilistic classification to each transient. In the second step, once the flight software is triggered, a subset of the event data is transmitted to ground-based 
centres, where further analysis and classification are conducted on the ground. The ground based analysis can take tens of seconds to a few minutes to complete \citep{thompson2024Fermi}. In the era of multi-messenger astronomy, where transient events are often unpredictable and short-lived, rapid and accurate identification is critical to enable timely 
follow-up observations. Such swift classification facilitates coordinated observations across multiple observatories and wavelengths, as well as through different messenger channels like 
neutrinos and gravitational waves. Therefore, there is a pressing 
need for alternative, faster, and more efficient approaches to transient classification that can complement or enhance existing systems.

With the increasing availability of data, machine learning, particularly its subfield - deep learning, has gained substantial 
significance and application in astrophysics \citep{mccullogh:logical, Kembhavi_Pattnaik_2022, Baron2019}. The rapid advancement of Graphics Processing Unit (GPU) technology has 
substantially increased computational capacity, enabling the training of complex models on large datasets, which are now common in modern astronomical surveys.
Deep learning algorithms, inspired by the architecture of the human brain, employ artificial neural networks (ANNs; \citealt{Zhang_2000}) 
to capture non-linear patterns in data. By stacking multiple interconnected layers, these models automatically extract relevant features from raw inputs, minimising the need for manual feature 
engineering typically required in conventional machine learning approaches.

Two of the most commonly used deep learning architectures are convolutional neural networks (CNNs; \citealt{lecun2015deep}) and 
recurrent neural networks (RNNs), each offering distinct advantages depending on the nature of the data. CNNs are particularly suited for extracting local and spatial features and have been widely used in 
astronomical tasks involving image-based data, including the classification of objects in the Optical Gravitational Lensing Experiment (OGLE; \citealt{Monsalves2024}), galaxy morphology prediction \citep{Dieleman2015}, transient detection in sky surveys 
\citep{Cabrera-Vives_2017}, and star-galaxy classification \citep{Bertin1996}.

RNNs, on the other hand, are designed to handle sequential data and excel in capturing temporal dependencies, making them especially 
effective for time-series applications where the order of data points holds crucial information. They have been applied in astrophysics for tasks such as redshift estimation \citep{10.1093/mnras/stae2446}, 
solar flare prediction \citep{Liu_2019}, and GRB light curve reconstruction \citep{Manchanda_2025}.

In this study, we developed and evaluated two deep learning models, based on combination of CNNs and RNNs architectures but differing in their structural design and approach, for multi-
class transient classification using {\it Fermi}-GBM data. Both models use multivariate time-series inputs derived from transient light curves 
binned at different time resolutions. The first model integrates convolutional and recurrent layers sequentially, while the second model 
employs multiple parallel convolutional branches for each time binning resolution, followed by a shared recurrent layer to capture temporal dependencies. Our approach builds upon prior 
studies such as \citealt{Abraham2021}, which demonstrated that transient classes exhibit distinctive light curve morphologies(Figure \ref{fig:lc_examples}) that 
can be exploited for classification, and \citealt{PengZhang2024} used CNNs for binary GRB/non-GRB classification based on count maps. While works like \citealt{Dimple_etal_2024} have employed Uniform Manifold Approximation and Projection with Principal
Component Analysis to cluster GRBs using their lightcurves. 

Expanding on this foundation, we apply Recurrent-Convolutional Neural Networks (R-CNNs) for the first time to a four-class problem involving GRBs, TGFs, SGRs and SFLAREs, achieving high classification accuracy with computational efficiency.
The CNN models are particularly effective in capturing local features and characteristic temporal patterns within short time bins, enabling fast inference. In contrast, RNN models leverage the sequential 
nature of time-series data to capture long-term dependencies and evolution patterns within light curves. By using these approaches in two 
different combined fashions, we systematically explore their individual strengths and demonstrate their complementary capabilities in accurately 
classifying high-energy transients, as well as in identifying potential outlier events indicative of rare or unknown phenomena.

The paper is organised as follows: in Section 2, we describe the data collection process and pre-processing steps. Section 3 details the 
methodology and the architectures employed in the two R-CNN model approaches. The results of our analysis are presented in Section 4. In section 5, we discuss about the misclassified events and the model limitations, the relevance of our work in the context of 
previous studies, its applicability to real-time classification, and the treatment of uncertain or outlier events with potential discovery prospects. Finally, section 6 provides a summary of our findings.

\section{Data}
This section outlines the procedures for data collection, preprocessing, and dataset balancing to ensure optimal model performance. 

\subsection{Data Collection}
\label{data_collection}

\replaced{}{The data used in this study is obtained from the {\it Fermi} Gamma-Ray Space Telescope, operated by NASA and publicly available through the High Energy Astrophysics Science Archive Research Center (HEASARC). }
{\it Fermi} is designed to detect gamma-ray bursts (GRBs) and other 
high-energy hard X-ray and gamma ray astrophysical phenomena. It carries two primary instruments: the Gamma-ray Burst Monitor (GBM), which includes 12 Sodium Iodide (NaI) detectors covering $8$ keV to $900$ keV and $2$ Bismuth Germanate (BGO) detectors covering $250$ keV to $40$ MeV, and the Large Area Telescope (LAT), which operates in the higher energy range of $30$ MeV to $300$ GeV. In this work, we use only GBM data, as it is responsible for triggering event detections, while LAT emission is observed for only a small fraction of cases.

The NaI detectors on GBM are oriented such as to provide near-complete sky coverage, excluding only the 
Earth-occulted region. Beyond detection, GBM also provides spectral information and real-time localisation of these transients. This 
rapid localisation enables re-pointing of the LAT detector and facilitates coordinated multi-wavelength follow-up observations by ground-based observatories, communicated through GCN Circulars.

The Data Processing Unit (DPU) onboard the GBM produces three types of data packets: CTIME, CSPEC, and TTE \citep{Meegan_2009}. For this study, only the TTE (Time-Tagged Event) files were used, specifically those from the top brighter NaI 
detectors (highest number of photon counts) with an angular separation of less than $60^{\circ}$ from the burst location. The TTE data provides the highest spectral and temporal resolution available, 
offering 128 energy channels and a temporal resolution of 2 microseconds. This makes TTE data ideally suited for detailed time-series analysis of high-energy transient events. 

The TTE data used in this study is obtained from the {\it Fermi} GBM Trigger Catalog (FERMIGTRIG\footnote{\url{https://heasarc.gsfc.nasa.gov/w3browse/fermi/fermigtrig.html}}) publicly available through the High Energy Astrophysics 
Science Archive Research Center (HEASARC). The {\it Fermi} GBM data used in this study were accessed from the archive covering the period June 2008 till March 2024.

Although a substantially larger number of GRBs ($\sim 3000$), SFLARES ($\sim 1800$), and TGFs ($\sim 1390$) have been detected by Fermi-GBM, the total number of reported SGR events is considerably smaller ($\sim 620$). In addition, from the Fermi GBM Trigger Catalog, we 
selected only those triggers assigned a 100\% classification probability by the GBM flight software \citep{Meegan_2009}. This probability represents the confidence level of the onboard classification algorithm \citep{Meegan_2009}.
To prevent biasing the training algorithm toward classes with substantially larger sample sizes, we constructed a balanced dataset by randomly selecting approximately 1000 high-confidence (100\%) events per class. This choice keeps the sample sizes comparable and close to the maximum 
number available for SGR events with 100\% classification confidence ($\sim 500$). Such a balanced sampling strategy mitigates class imbalance during training and ensures that the overall dataset size remains compatible with the available computational resources.



\subsection{Data Preprocessing}


\begin{figure*}[t]
    \centering
    \includegraphics[width=1\textwidth]{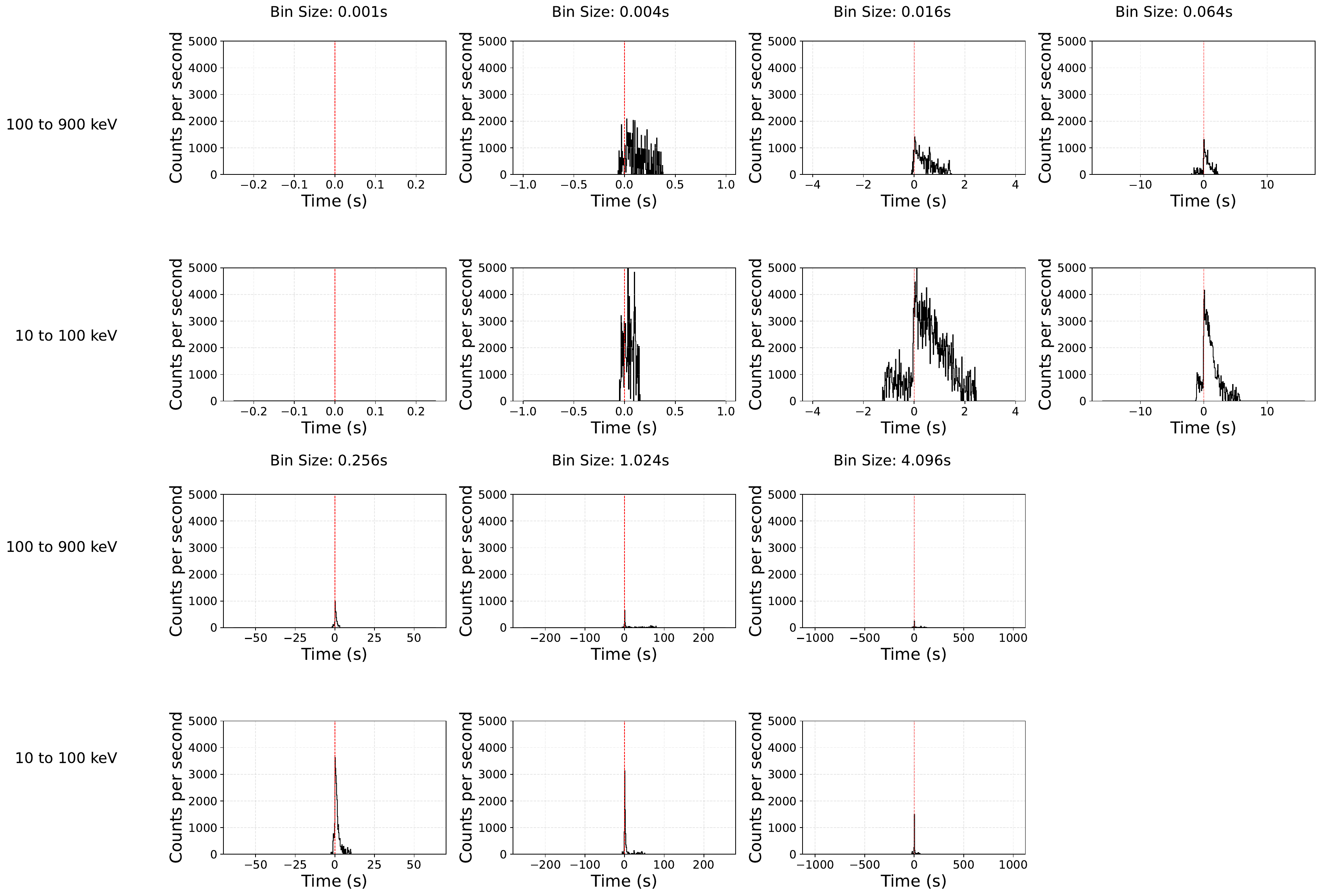}
    \caption{The background-subtracted light curves used as model inputs are shown above. Each sample comprises 14 light curves from two energy ranges (10–100 keV and 100–900 keV) and seven time-bin resolutions. When the algorithm fails to clearly separate signal from background, an empty light curve is passed to the model.}
    \label{fig:sublc_example}
\end{figure*}

The TTE data primarily contain two key pieces of information: the detection time and the energy of each photon. By binning the photon detection times from the TTE file, a counts per second versus time plot is generated. This is referred to as the light curve of the transient. Since the light curve captures the temporal evolution of photon counts, it is treated as time series data.

The typical durations of high-energy transients span a wide range, from sub-millisecond for TGFs \citep{Mailyan_etal_2020} and about 0.01–1 s for SGRs \citep{Gogus_etal_2001}, to roughly 0.1–100 s for GRBs \citep{Fermi_GRB_catalog_2020}, and several minutes to hours for SFLAREs \citep{Reep_etal_2019}.
Therefore, the detectability of transients in light curve analysis is governed by their intrinsic timescales, as different transient durations achieve optimal signal-to-noise ratios (SNR; $>4 \sigma$ in given energy ranges; \citealt{Fermi_GRB_catalog_2020}) at different time 
binnings (Figure \ref{fig:sublc_example}). To account for this variability, a single binning scheme was insufficient to capture the diversity of the transient signals. A detailed analysis of light curves from various transient classes 
allowed the identification of seven optimal bin sizes - 0.001 s, 0.004 s, 0.016 s, 0.064 s, 0.256 s, 1.024 s, and 4.096 s - chosen to maximise the SNR across different event types.

\begin{figure*}[t]
    \centering
    \includegraphics[width=1\textwidth]{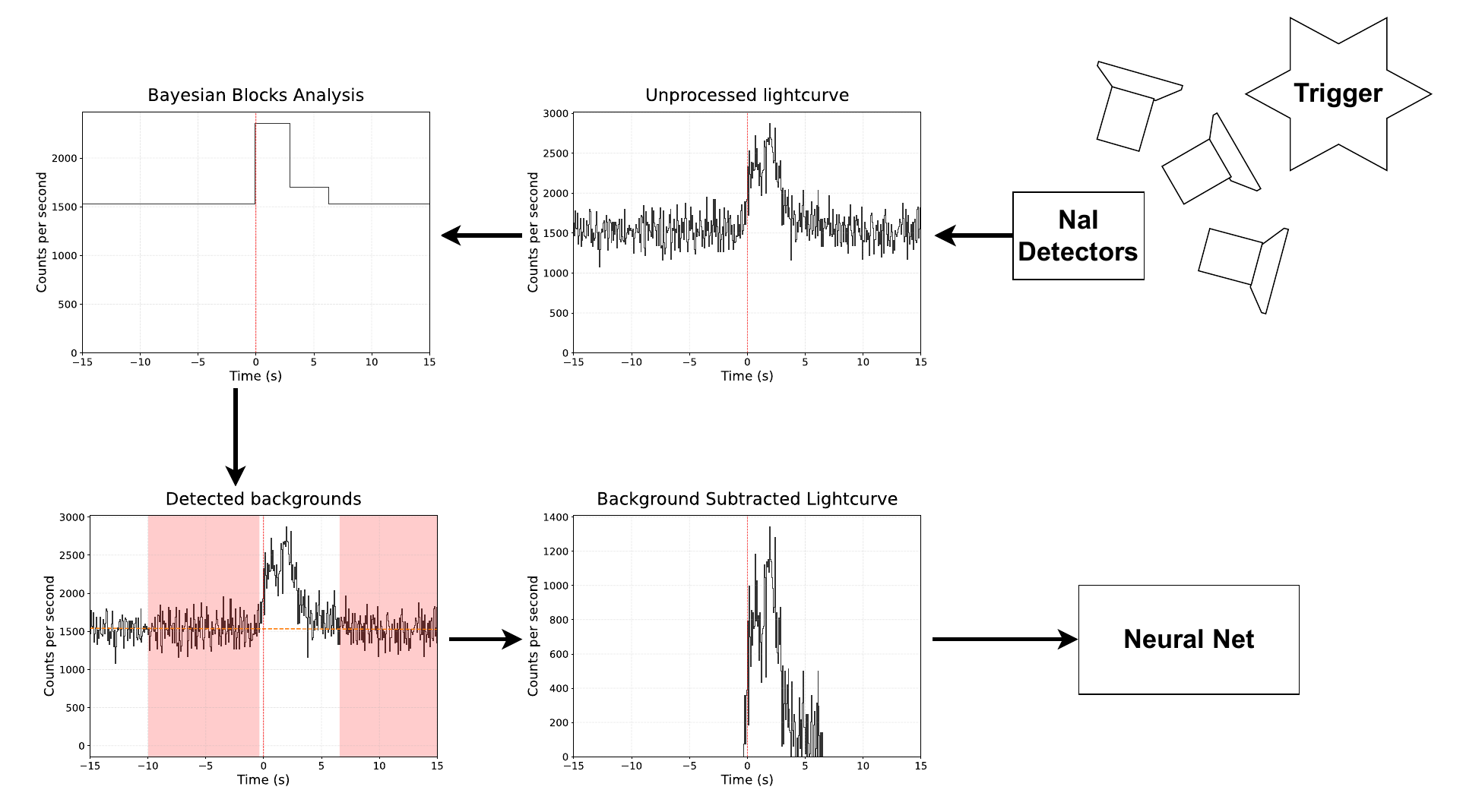}
    \caption{Schematic illustration of the data processing pipeline, showing the complete workflow from data extraction to background removal and generation of clean data arrays.}
    \label{fig:data_processing}
\end{figure*}

For each trigger, the three brightest TTE files from the NaI detectors closest to the source direction were selected and combined to create light curves in two energy ranges: 10--100 keV (channels 3--50) and 100--900 keV (channels 51--124) (Figure \ref{fig:sublc_example}). 
The light curves were constructed using 500 bins centered at the trigger time (set to zero), excluding partially filled bins and padding the edges with zeros to ensure uniform length. The Bayesian block algorithm 
was then applied to identify the transient start and stop times, as well as the pre- and post-burst background intervals. The background was modeled and subtracted, and 
light curves without a statistically significant signal were discarded to prevent the model from learning noise. Details of these steps are provided in Appendix \ref{appendix:data_processing}, and a schematic overview of the procedure is shown in Figure \ref{fig:data_processing}.
\replaced{}{The photon arrival times\footnote{The arrival times of the photons and the trigger time of the burst are reported in Mission Elapsed Time (MET) which for {\it Fermi} is the number of seconds since 00:00:00 UTC on January 1, 2001 (UTC).} 
were first adjusted by subtracting the trigger time, effectively setting the trigger time to zero.}
\deleted{For each of the seven bin sizes, the bin edges were calculated using the following equations:
where $dno = 500$ is the total number of bin edges, $r$ is the ratio 
of pre-trigger to post-trigger bins, and $i$ is the bin size. 
When the counts in the TTE files did not 
fully span the defined time range, some bins were only partially populated. To ensure consistency, such bins were excluded, and the light curves were padded by replacing the non-zero counts at the extremes with zeros, effectively limiting each light curve to 500 bins centered around the trigger time.}
\replaced{}{
\replaced{Following this, the Bayesian block algorithm}{The Bayesian block algorithm} \citep{scargle2013studies} with a false alarm probability ($p_0 = 0.01$) was applied to each light curve 
using the Python package \texttt{pwkit} (version 1.2.0) 
\citep{williams2017pwkit} 
to extract significant signal features and 
allowing to identify the start and stop times of each transient, as well as facilitated the identification of pre- and post-transient 
background regions in each light curve. The background was then modeled and subtracted from the signal.}
\replaced{}{ The schematic diagram depicting the data processing is shown in Figure \ref{fig:data_processing}.}
\replaced{}{
\replaced{If for any reason the algorithm failed to identify a significant signal in any particular energy or time binning, the data from that binning was omitted to prevent the model from learning from noise.}{We further note that if the Bayesian block analysis detected only two time blocks, or if the interval between the extreme blocks (representing the background regions) was too short relative to the bin size, the event was 
identified as lacking a significant signal for that particular binning and energy range. If this occurs we send an array with zero signal to prevent the network learning from data without significant signal in that specific time binning and energy range.}} 

\deleted{The background was modeled using a polynomial fit (up to the fourth order was investigated) in the identified pre and post transient time intervals. {The best fit was identified using reduced chi-square closest to 1.} Subsequently, the polynomial was subtracted from the lightcurve. This procedure was repeated across all seven binnings and both energy ranges.}
As a result, for each trigger, a total of 14 background-subtracted light curves were generated (seven bin sizes $\times$ two 
energy bands). Each light curve was formatted as a time series with two feature channels (representing the two energy ranges), resulting in an array shape of $(\text{samples}, 499, 2)$. \deleted{, where the number of samples varied depending on the specific dataset under analysis.} This standardized multi-resolution and multi-energy representation ensured 
that both temporal structures and spectral hardness information were preserved for input into the machine learning models.

For approximately $5\%$ of the triggers in our initial dataset, this method fails to detect a signal in any of the studied time binning or energy intervals. These triggers are therefore excluded from the training and test sets (Table \ref{tab:data}). The characteristics of these events are discussed in Section \ref{sec:eff}.


\begin{table}[!htp]\centering
 \scriptsize
\begin{tabular}{lcccc}
\hline
\textbf{Partition} & \textbf{No. of GRBs} & \textbf{No. of TGFs} & \textbf{No. of SGRs} & \textbf{No. of SFLAREs} \\
\hline
Total & 990 & 866 & 477 & 974 \\
Train & 790 & 691 & 387 & 776 \\
Validation (20\% Train) & 158 & 138 & 77 & 155  \\
Test & 200 & 175 & 90 & 198 \\
\hline
\end{tabular}
\caption{The final size of data set used for training, validation and testing the models are give above.}
\label{tab:data}
\end{table}

For the first deep learning model, R-ConvNet, the histogram counts from different bin sizes were stacked into a single array. These arrays were then concatenated 
to form a three-dimensional input of shape (samples, channels, time steps), where each channel corresponds to a specific bin size. The model was trained 
using inputs of shape (channels, time steps), allowing it to capture variability and patterns across multiple temporal resolutions for each event. 

In the second deep learning, MI-DCL (Multi Input - Dual Channel Lightcurve), a similar approach was followed; however, instead of stacking the histogram counts into a single array, each bin size was treated as a separate input. Seven distinct input layers corresponding 
to the seven bin sizes were fed into the model and merged into a single tensor via concatenation. This design enabled the model to process each time 
resolution independently before combining the extracted features. 

\deleted{A holdout strategy was employed to split the dataset into training, validation, and test sets. The identified dataset (as mentioned in section 
\ref{data_collection}) was first divided into training and test sets, with a portion of the training set further set aside for validation during model 
training. The model was trained on the remaining training data, while the validation set was used to monitor performance after each epoch and to fine-
tune hyperparameters. Finally, once training was complete, the model’s performance was evaluated on the test set, which remained unseen during both training and validation.}

\subsection{Data Balancing and Splitting}
A holdout strategy was adopted to divide the identified dataset (Section \ref{data_collection}) into training and test sets, with the test set kept completely unseen 
during model development. From the training set, 20\% was randomly split and used as a validation subset during training in TensorFlow.

The model was trained on the remaining 80\% of the training data, while the validation subset (20\% of training set) was used exclusively to monitor performance 
after each epoch and to guide hyperparameter tuning. No independent validation dataset was maintained. Final model performance was evaluated only on the held-out test set.

Approximately 800 samples each of GRBs, TGFs, and SFLAREs, and 400 samples of SGRs were assigned to the training set, of which 20\% 
was allocated for validation. The test set comprised roughly 200 samples each of GRBs, TGFs, and SFLAREs, and 92 samples of SGRs. The 
final dataset sizes used for training, validation, and testing are summarised in Table \ref{tab:data}. This procedure ensured a balanced class distribution while preventing data leakage.



\section{Methodology} \label{sec:methods}

Both our classifiers were built using widely adopted deep learning architectures: Convolutional Neural Networks (CNNs) and Recurrent Neural Networks (RNNs). CNNs \citep{lecun2015deep} are adept at learning spatial patterns and are particularly useful in identifying both local and hierarchical features. For example, in the case of 1D time-series data such as transient light curves, CNNs treat the temporal axis as a spatial dimension, allowing the network to recognise features like sudden spikes or spectral hardness variations. Each subsequent layer in the CNN hierarchy learns increasingly abstract representations, making CNNs highly effective in extracting complex patterns while filtering out noise.


RNNs (\citealt{info15090517}) are specifically designed for sequential and time-series data. Unlike standard ANNs, where each input is processed independently, RNNs are capable of retaining context from previous inputs, making them well-suited for modeling sequences where current outputs depend on prior states. This is achieved by introducing a hidden state that carries information forward through the sequence.




In this study, we utilise Long Short-Term Memory (LSTM) networks, a specialised form of RNN proposed by \cite{10.1162/neco.1997.9.8.1735}, designed to overcome challenges such as vanishing 
gradients in long sequences. LSTM networks introduce memory cells regulated by three gates - input, forget, and output gates - that determine which information to retain, discard, or output at each step. This mechanism enables LSTMs to capture 
long-term dependencies within sequences effectively. Given that our dataset consists of time-series photon count data, LSTMs are well-suited to capturing hidden temporal patterns essential for accurate transient classification.

\subsection{Neural Network Architecture}

The two R-CNN models developed by our team have distinct architectures. The main distinguishing feature of the two R-CNNs was how they handled the heterogeneous nature of the data due to different bin sizes. 

\subsubsection{R-ConvNet}

In the first model, the data is first passed through a convolutional input layer. This layer applies a set of 
filters (kernels) to extract low-level features from the input data. Following the initial convolution, a max-pooling layer is employed to 
reduce the length of the temporal dimension of the resulting feature maps. This dimensionality reduction compresses the representation, decreases the number of learnable parameters, and helps mitigate overfitting.

Subsequently, an additional convolutional layer is added, followed by an LSTM (Long Short-Term Memory) layer, with a 
max-pooling layer in between. This architecture allows the model to capture both spatial and temporal dependencies in the data—where convolutional layers focus on extracting 
local features, and LSTM layers capture sequential patterns. After these layers, a global max-pooling layer 
is applied, which operates similarly to standard max pooling but aggregates over the entire time axis, enabling 
the model to handle variable-length inputs.

The pooled output is then flattened and passed through a series of fully connected dense layers, which further 
learn high-level representations and patterns from the extracted features. Except for the output layer, each 
layer uses the ReLU (Rectified Linear Unit) activation function \citep{nair2010rectified}, which is computationally efficient and helps mitigate the vanishing 
gradient problem. 

Following the dense layers, three fully connected layers are included with dropout layers placed in between to prevent overfitting by randomly deactivating a fraction of neurons during training \citep{JMLR:v15:srivastava14a}. The final output layer uses a softmax activation function, appropriate for multi-class classification tasks, which converts the model outputs into probability distributions over the classes. The softmax function is given by:
\begin{equation}
\textit{Softmax}(z_i) = \frac{e^{z_i}}{\sum_{j=1}^{C} e^{z_j}},
\label{eqn:soft}
\end{equation}
where $C$ is the number of classes and $z_i$ is the logit (input) to the softmax for class $i$.

\begin{figure}[htbp]
    \centering
    \includegraphics[width=1\linewidth]{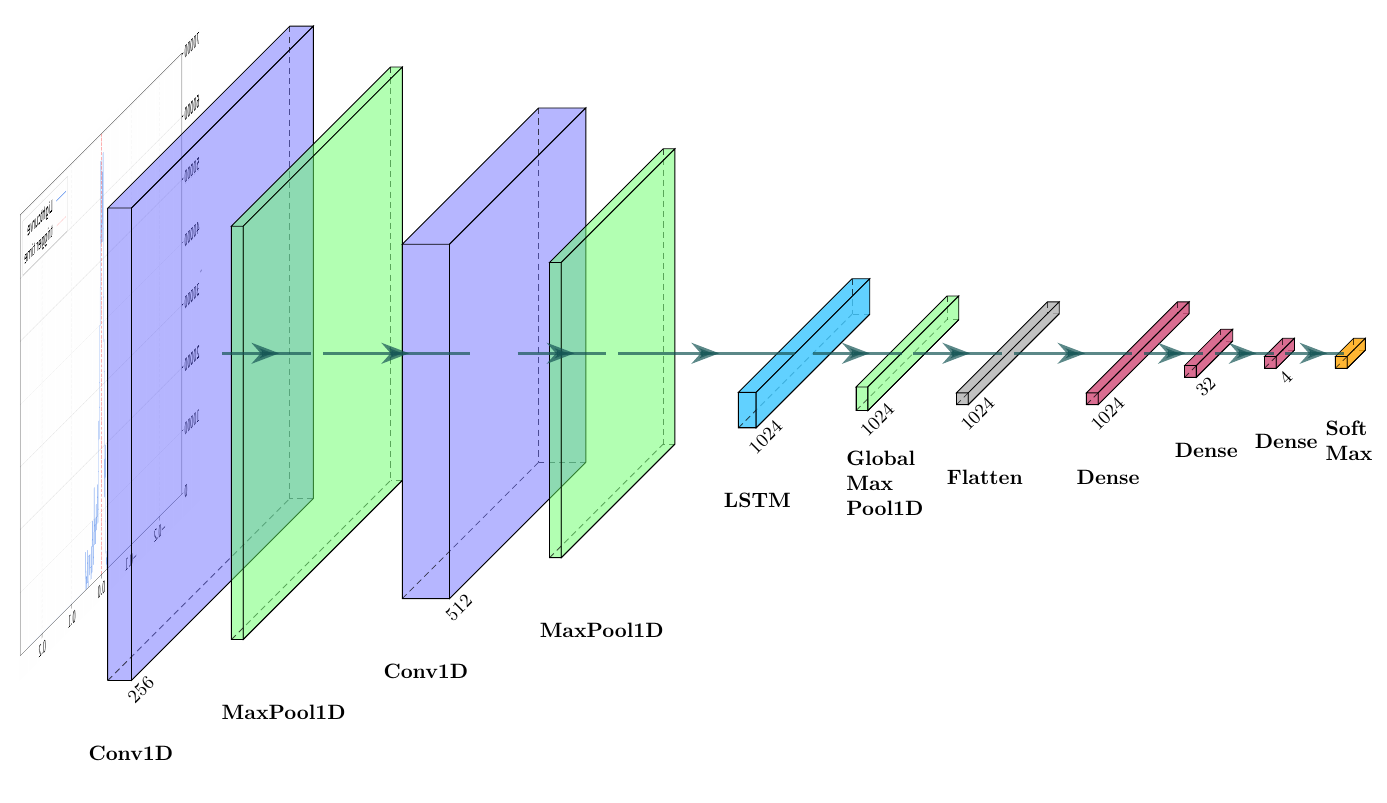}
    \caption{The schematic diagram of the proposed R-ConvNet model illustrates the detailed architecture, showing how the input data are processed through successive layers to learn underlying spatial and temporal features.}
    \label{fig:R-ConvNet}
\end{figure}

Additionally, L2 regularisation is applied to all convolutional and dense layers to further prevent 
overfitting by penalising large weight values, which also stabilises the loss curve. Model training is performed using the Adam (Adaptive Moment Estimation) optimiser 
\citep{DBLP:journals/corr/KingmaB14}, which adaptively adjusts the learning rate during training. The loss function used is categorical cross-entropy, standard for multi-class classification
This loss function quantifies the difference between the predicted and true distributions, guiding the model to make accurate predictions. The default \textit{EarlyStopping} and \textit{ReduceLROnPlateau} callbacks in keras are integrated to monitor the validation loss. \textit{EarlyStopping} stops training if the loss doesn't improve for a fixed number of epochs, while \textit{ReduceLROnPlateau} reduces the learning rate under same condition. The schematic representation of the model architecture is shown in Figure \ref{fig:R-ConvNet}.

\begin{figure}[ht]
    \centering
    \includegraphics[width=1\linewidth]{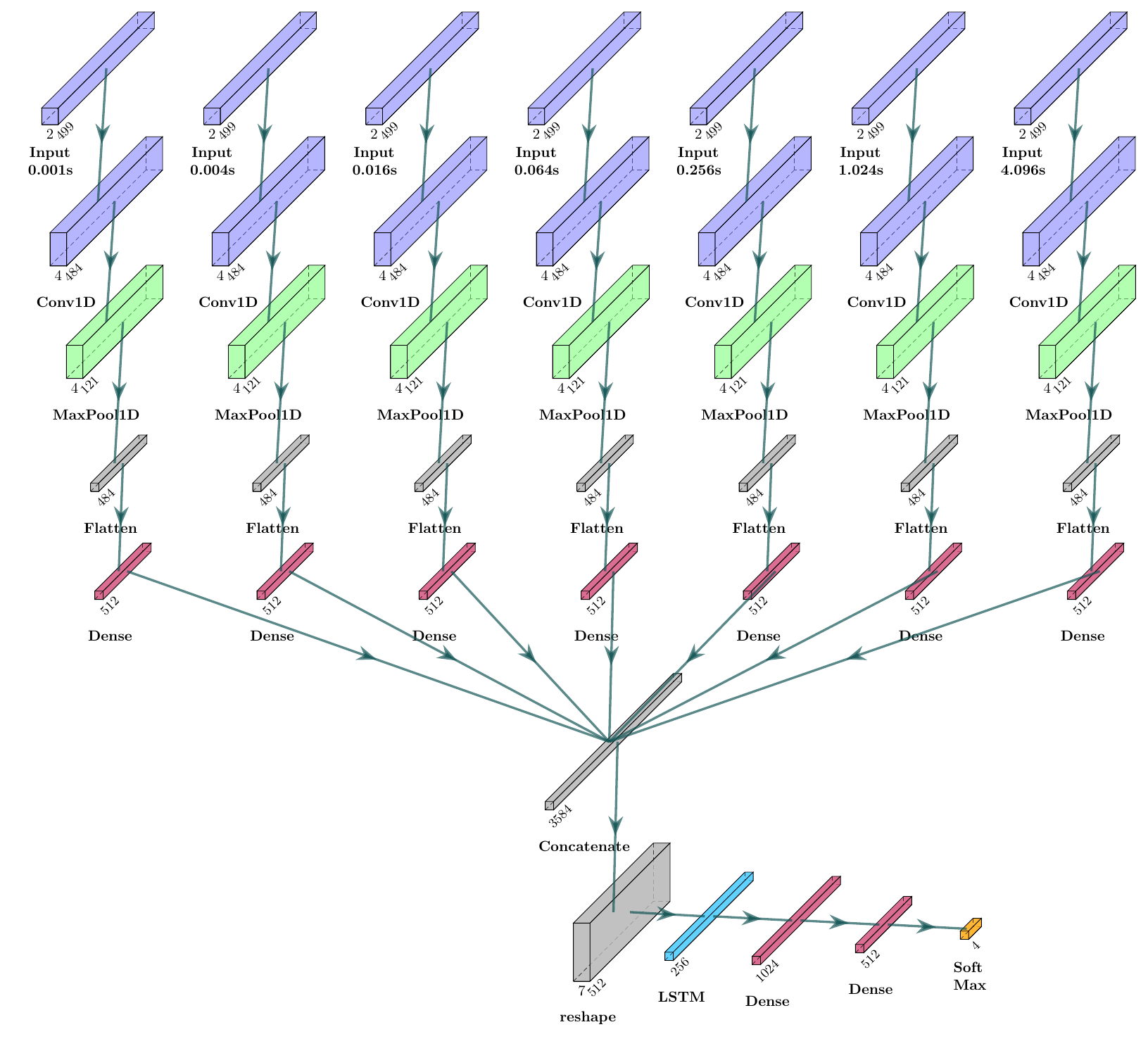}
    \caption{The schematic diagram of the proposed MI-DCL model illustrates the detailed architecture and how the input data are processed through successive layers to learn underlying spatial and temporal features.}
    \label{fig:MIDCL-model}
\end{figure}

\subsubsection{Multi-Input Dual Channel Lightcurve (MI-DCL) model}

The second R-CNN model, referred to as the Multi-Input Dual Channel Lightcurve (MI-DCL) model, employs separate input 
branches for each bin size. Each of the seven light curve inputs is processed through an independent pathway. The input format follows the time series structure of the 
light curves, with two features representing photon counts in the two energy bands

Each input branch consists of a one-dimensional convolutional layer with four filters of size 16, followed by a max-pooling layer with a pool size of 4. The output is then flattened and passed through a fully connected layer with 512 neurons using ReLU activation. 
as 
To mitigate overfitting, a dropout layer with a dropout rate of 0.25 
is applied. This structure is replicated independently across all seven light curve inputs.

The extracted features from all seven branches are then concatenated into a single merged feature vector. This 
vector is reshaped into a two-dimensional tensor and passed through a Long Short-Term Memory (LSTM) layer 
comprising 256 units, with both dropout and recurrent dropout rates set at 0.3 to further reduce overfitting and improve generalization.

Subsequent to the LSTM layer, the architecture includes a dense layer with 1024 neurons and ReLU activation, 
followed by a dropout layer with a 0.5 dropout rate. This is followed by another dense layer with 512 neurons and 
ReLU activation, again accompanied by a dropout layer with a rate of 0.5. Finally, a softmax output layer, as 
described in equation \ref{eqn:soft}, is used to produce class probabilities across four transient classes.

The model is optimised using the Adam optimiser with a learning rate of $0.0005$ and $\beta_1 = 0.9$ {(exponential decay rate for the first moment estimates)}. A learning rate scheduler is implemented to reduce the learning rate 
by a factor of $0.5$ if the validation loss plateaus for three consecutive epochs, with a minimum learning rate 
capped at $1 \times 10^{-6}$. Additionally, a custom early stopping mechanism is employed: training halts if the model achieves a training accuracy of $95\%$ or higher and maintains this performance for three consecutive epochs. This approach ensures efficient training, reducing unnecessary computation and mitigating overfitting risks.
The loss function used is categorical cross-entropy.

The schematic diagram of the model architecture is provided in Figure \ref{fig:MIDCL-model}. Overall, this architecture effectively combines convolutional feature extraction with temporal sequence modeling, enabling the model to capture both spectral and temporal characteristics of transient light curves. The MI-DCL model achieves robust multi-class classification performance across diverse transient types and was trained for a total of 88 epochs.

\section{Results}

\begin{figure}[ht]
    \centering
    \subfloat[R-ConvNet]{\includegraphics[width=0.45\textwidth]{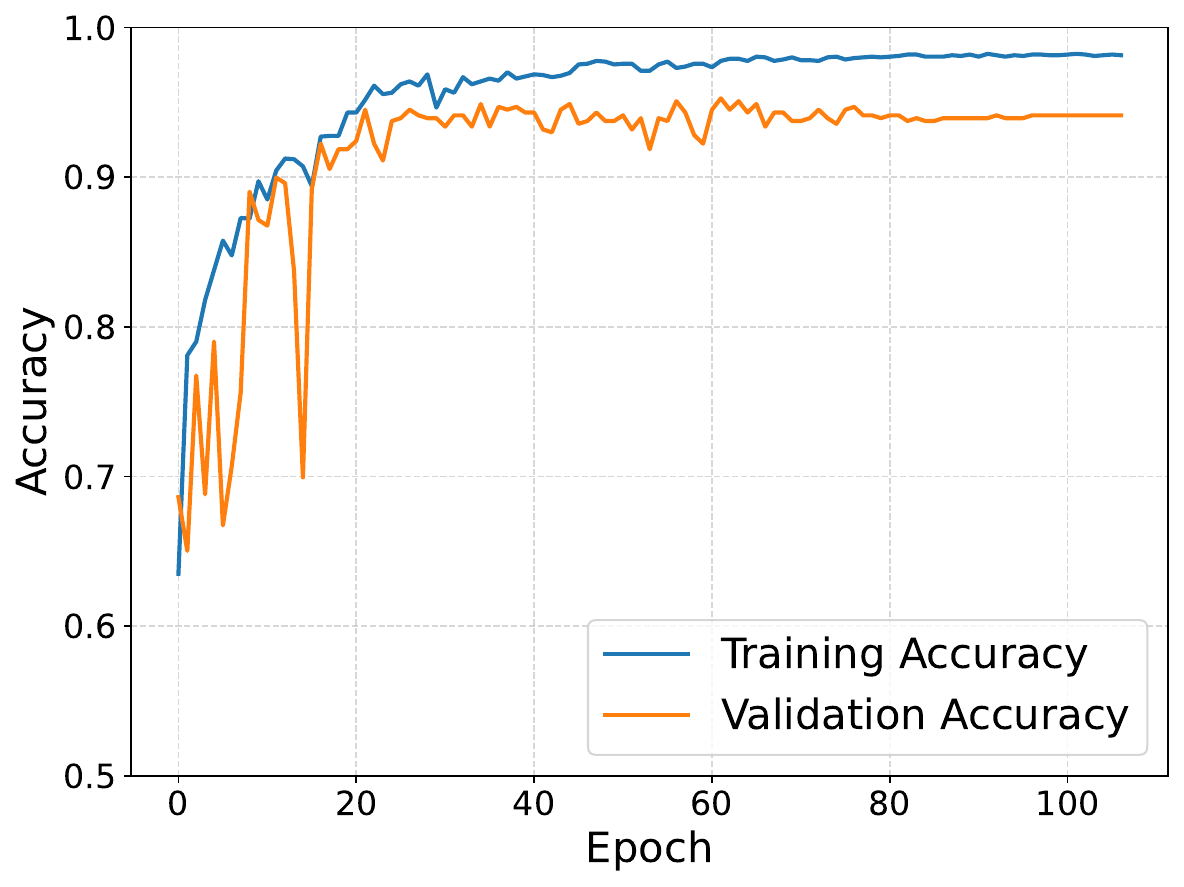}}\hfill
    \subfloat[R-ConvNet]{\includegraphics[width=0.45\textwidth]{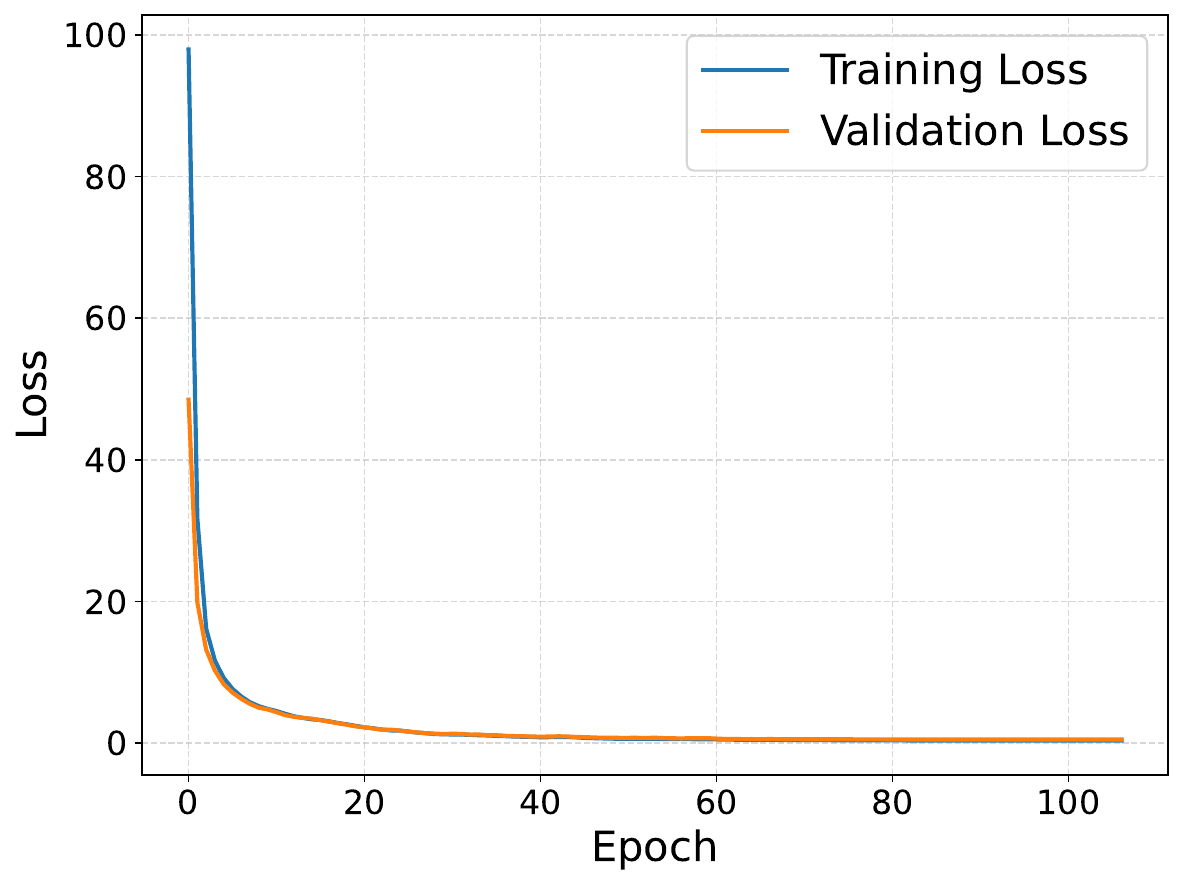}}\\
    \subfloat[MI-DCL]{\includegraphics[width=0.45\textwidth]{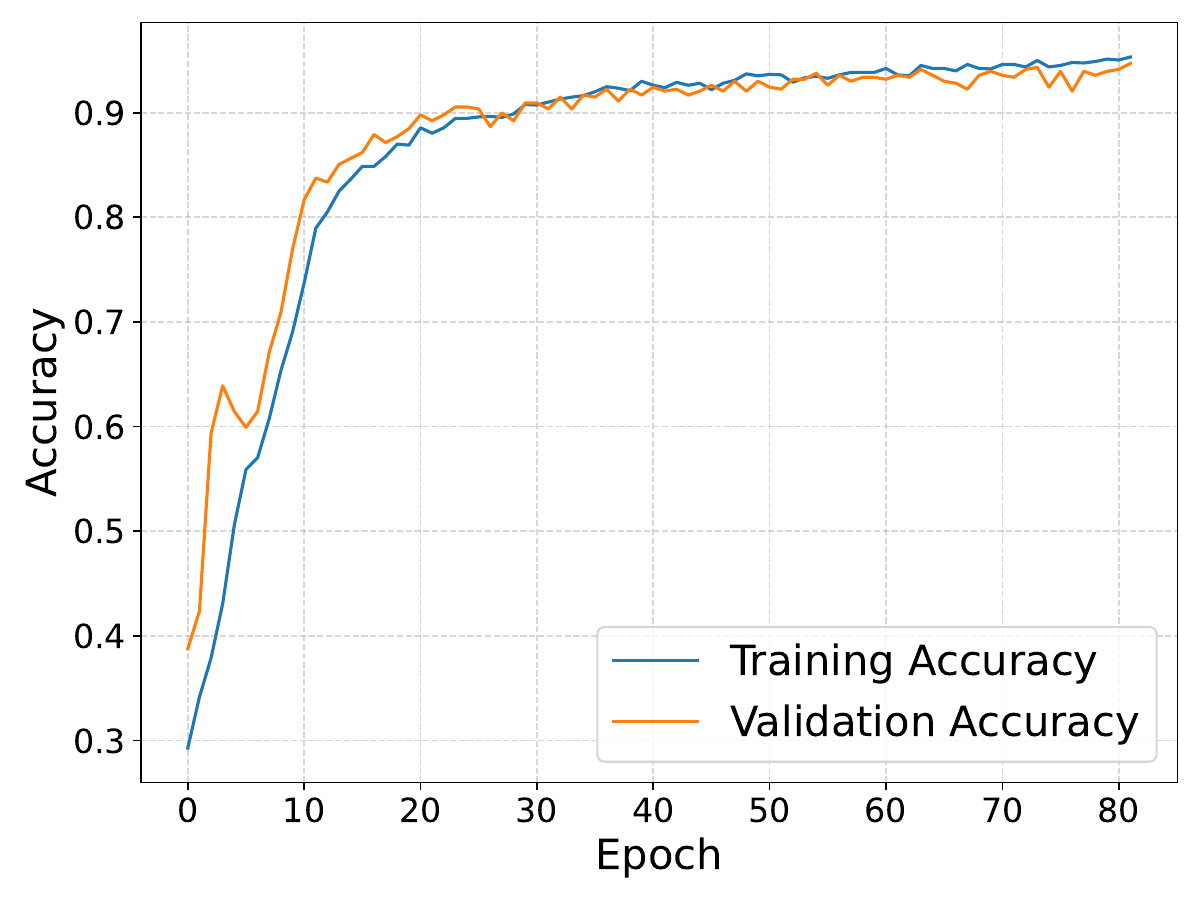}}\hfill
    \subfloat[MI-DCL]{\includegraphics[width=0.45\textwidth]{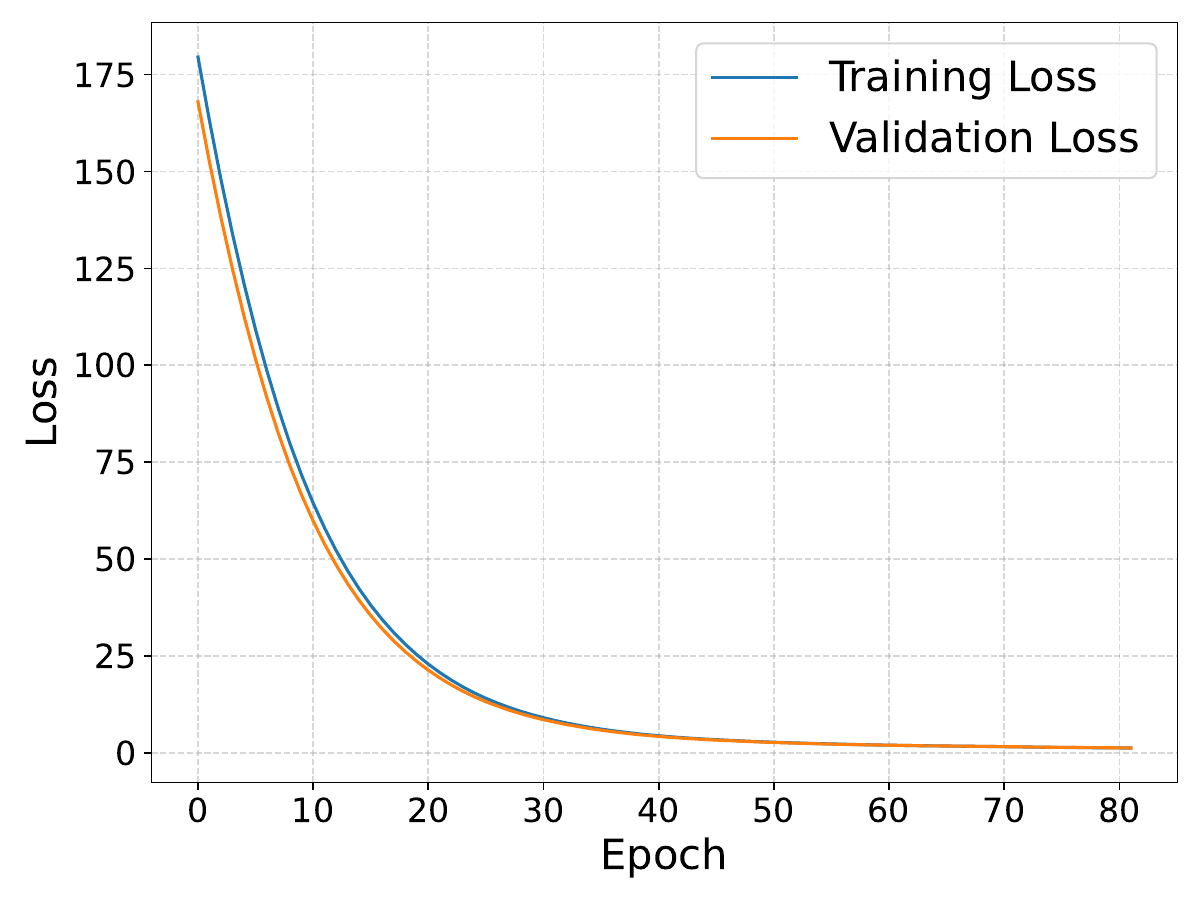}}
    \caption{The training performance of our deep learning models for transient classification is shown above. Panels (a) and (c) present the accuracy–epoch curves  while panels (b) and (d) display the corresponding loss–epoch curves for the R-ConvNet and MI-DCL models, respectively. The steady decrease and subsequent stabilization of both training and validation losses indicate that the models generalize well and are not prone to overfitting, despite the relatively limited dataset.}
    \label{fig:train_perf}
\end{figure}


\begin{figure}[ht]
    \centering
    \subfloat[R-ConvNet]{\includegraphics[width=0.45\textwidth]{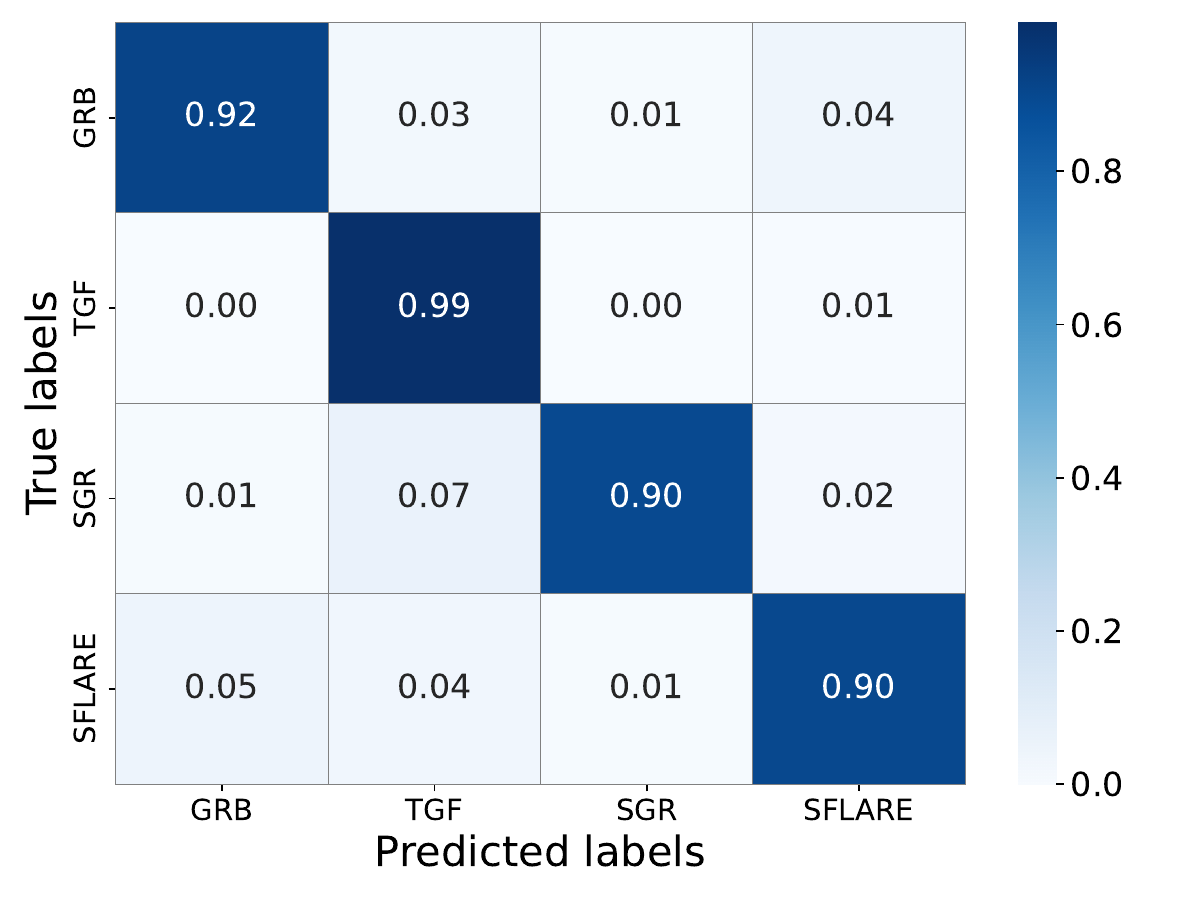}}\hfill
    \subfloat[R-ConvNet]{\includegraphics[width=0.45\textwidth]{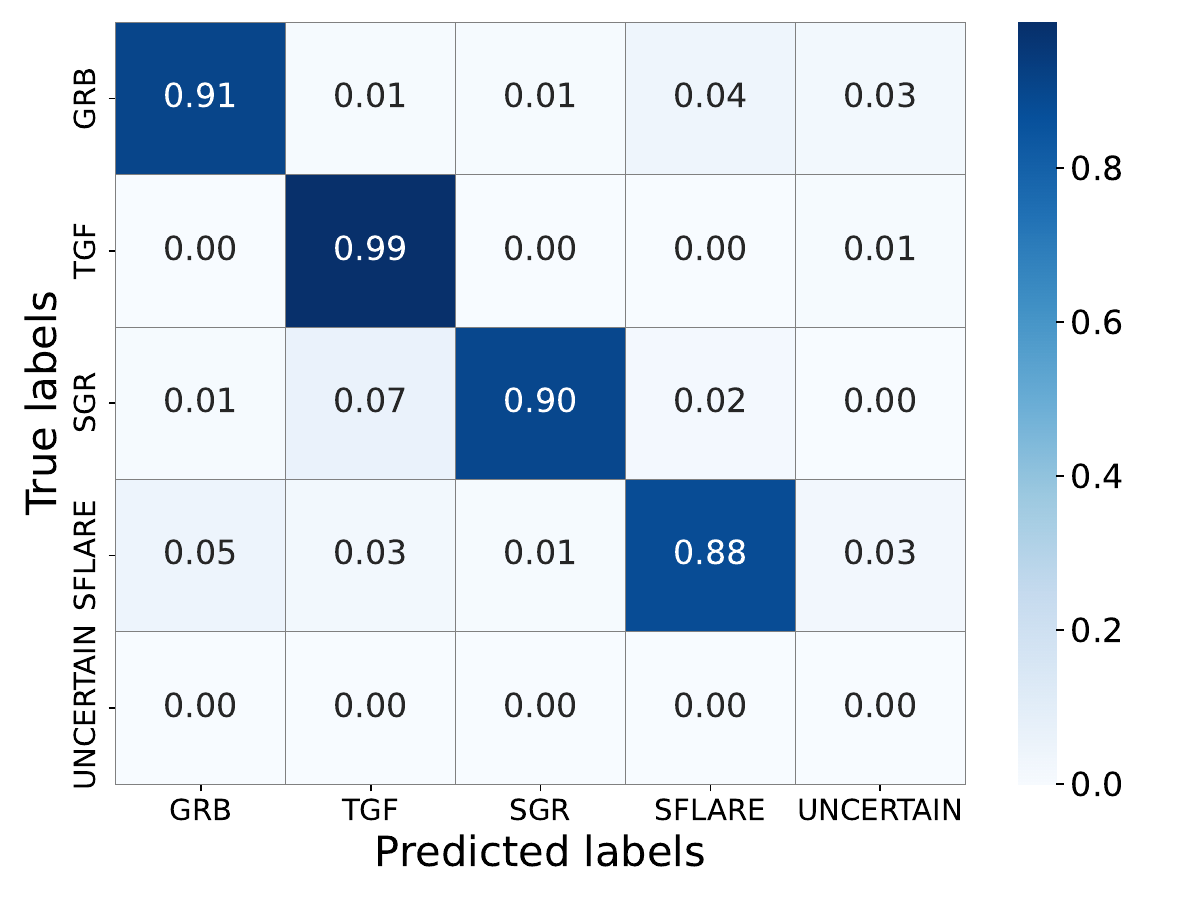}}\\
    \subfloat[MI-DCL]{\includegraphics[width=0.45\textwidth]{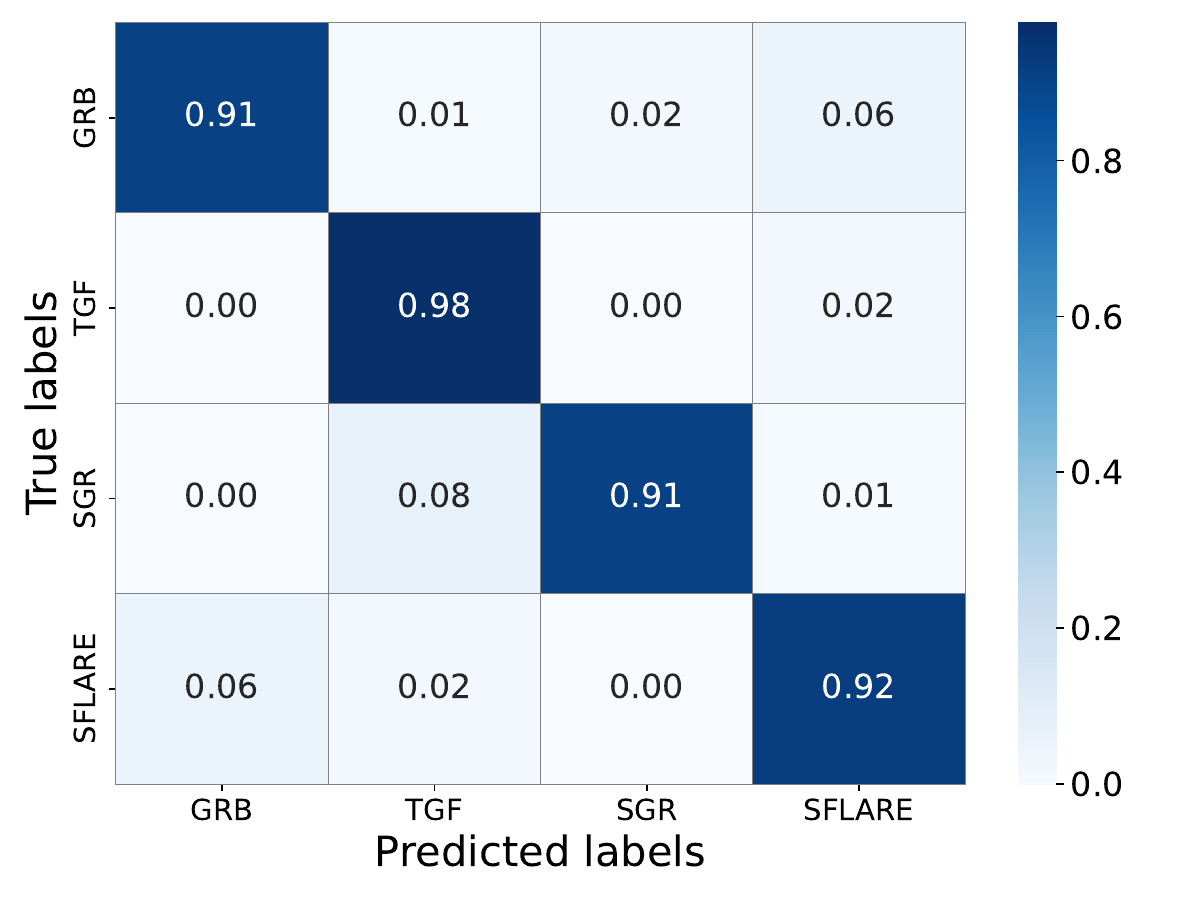}}\hfill
    \subfloat[MI-DCL]{\includegraphics[width=0.45\textwidth]{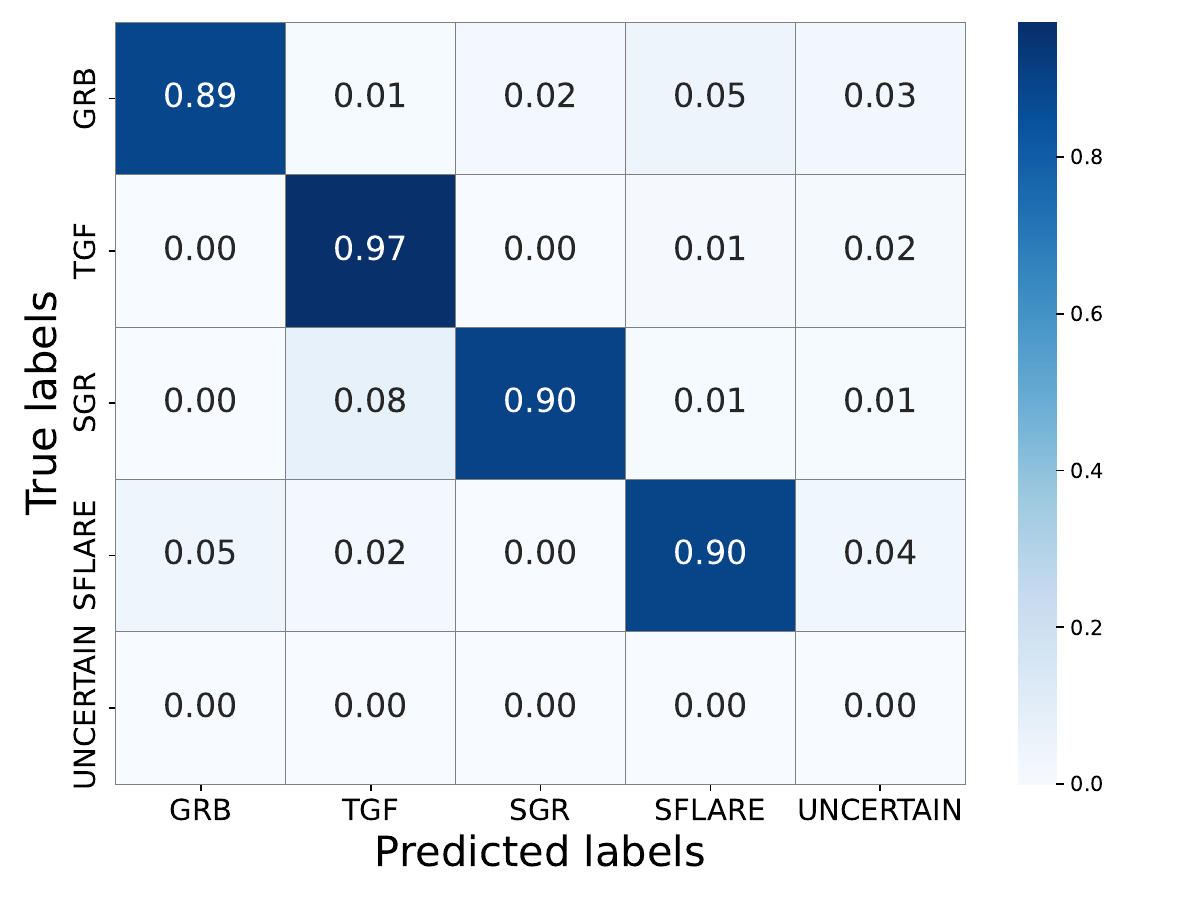}}
    \caption{Confusion matrices of both R-ConvNet [(a), (b)] and MI-DCL [(c), (d)] models on test data. (a) and (c) show the standard confusion matrices, displaying classification performance for GRBs, TGFs, SGRs, and SFLAREs. (b) and (d) include classifications with an additional 'UNCERT' category.}
    \label{fig:conf_mat}
\end{figure}

\begin{figure*}[ht]
    \centering
    \subfloat[R-ConvNet]{\includegraphics[width=0.45\textwidth]{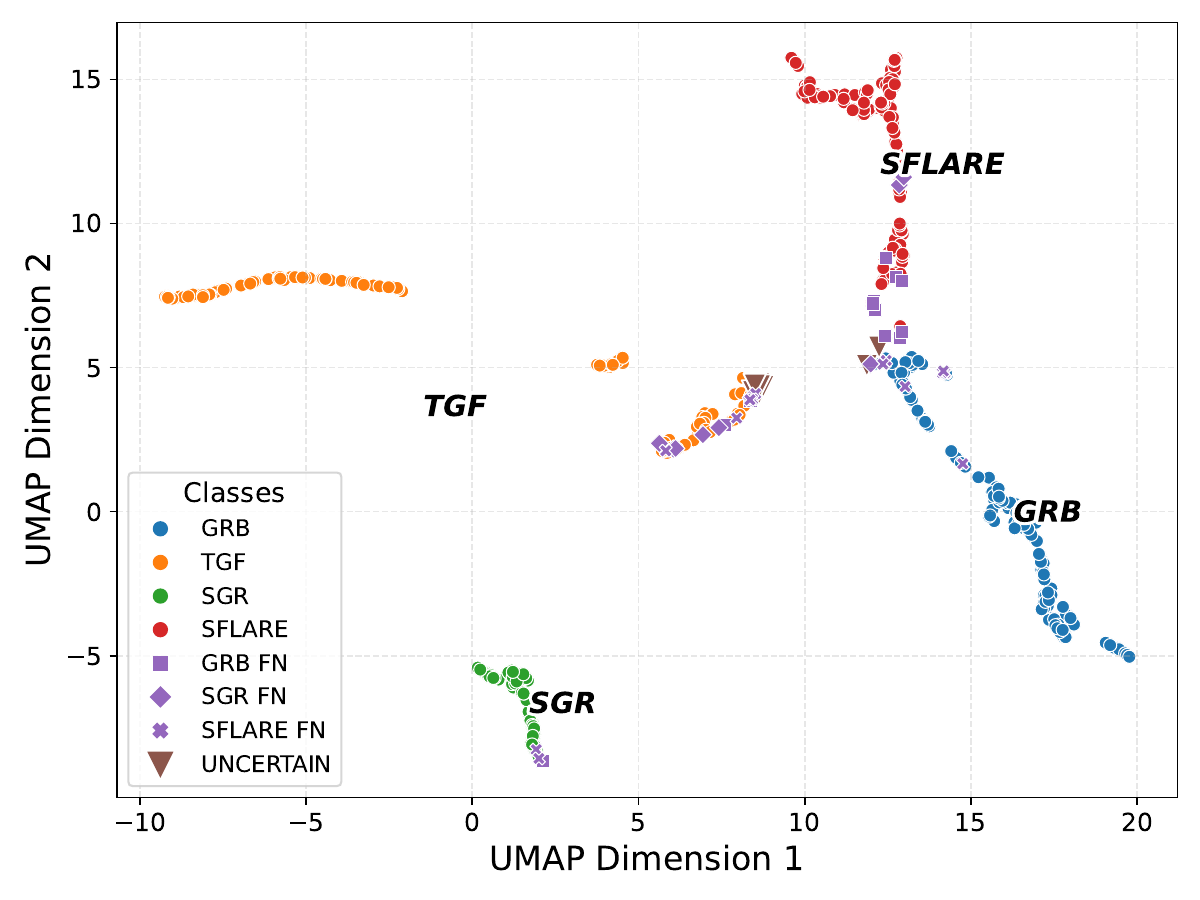}}\hfill
    \subfloat[MI-DCL]{\includegraphics[width=0.45\textwidth]{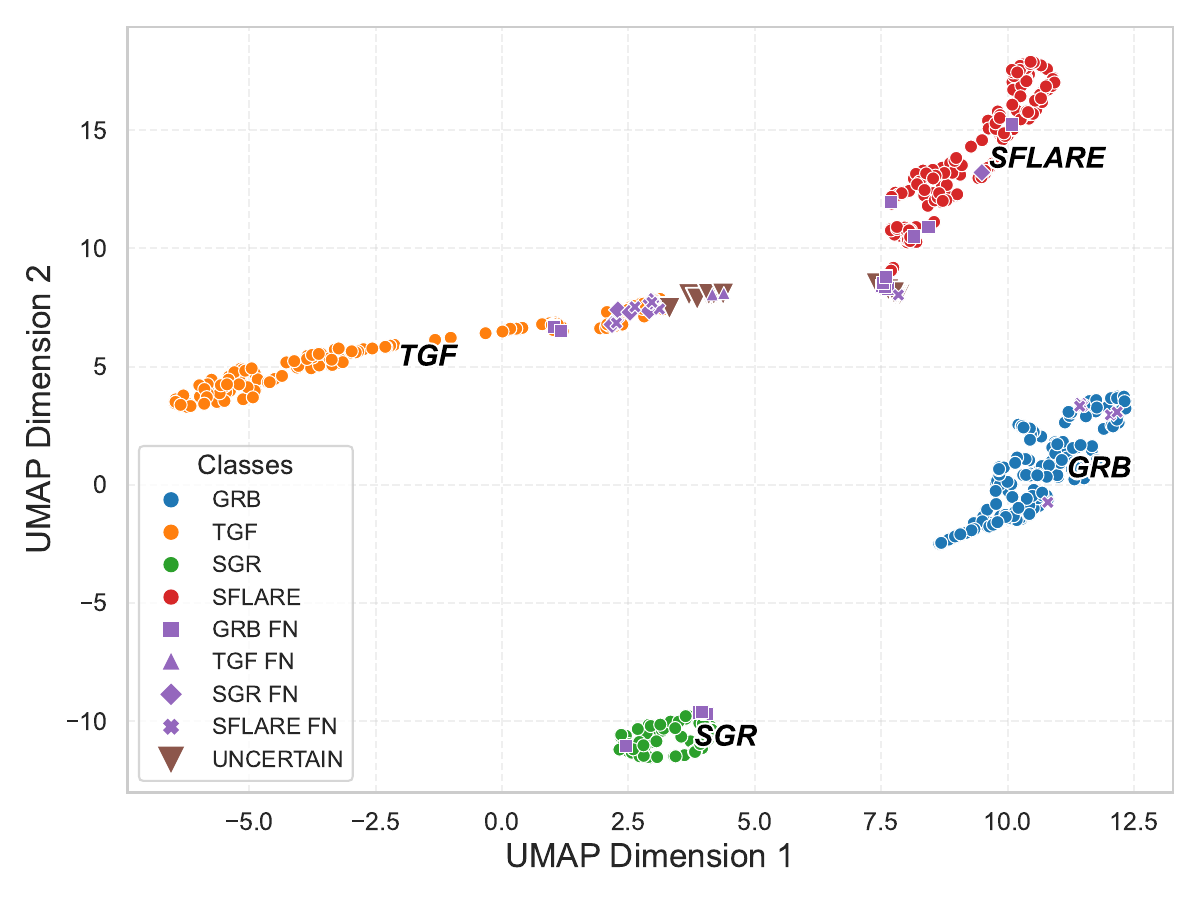}}
    \caption{Two-dimensional UMAP visualization of (a) R-ConvNet model and (b) MI-DCL model for multiclass classification of transients from our test set. 
    The learnt data representation was extracted from the penultimate hidden layers of each model. The plots display correctly classified transients (GRB - blue, TGF - orange, SGR - green, SFLARE - red), mislabeled events (purple), and uncertain events (brown). Each transient class forms distinct clusters, demonstrating the model’s ability to separate different classes effectively. As expected, uncertain events tend to lie near cluster boundaries, indicating regions of higher classification ambiguity.}
    \label{fig:umap}
\end{figure*}
 


Two R-CNN models were developed and evaluated using both built-in methods in Keras and custom-built functions. Model performance was assessed on the test set, with the learning curves presented in 
Figure~\ref{fig:train_perf}.

To provide a detailed evaluation of classification performance, we present the confusion matrices for both models in 
Figure~\ref{fig:conf_mat}. The confusion matrix summarises the number of correctly and incorrectly classified instances for each transient class, thereby forming the basis for 
all reported performance metrics. In particular, the diagonal elements represent correctly classified events, while off-diagonal elements indicate 
misclassifications between classes. Given the class imbalance in our dataset, particularly the under-representation of SGRs relative to GRBs, we report row-normalised confusion matrices in fractional form 
to enable a more meaningful class-wise comparison beyond overall accuracy. The actual number of events contributing to each entry can be inferred from the test-set sizes listed in Table~\ref{tab:data}.

From the confusion matrix, we compute four standard evaluation metrics: Accuracy, Precision, Recall, and F1-score. 
Accuracy represents 
the overall fraction of correctly classified instances. However, because the dataset is imbalanced, 
accuracy alone may not fully capture model performance. Precision measures the fraction of correct 
positive predictions among all predicted positives, while Recall quantifies the 
fraction of true positives correctly identified. The F1-score, defined as the harmonic mean of Precision and Recall, provides a balanced metric that is especially informative under class imbalance. The resulting classification reports are summarised in Table~\ref{tab:class_report}. Both models achieve an overall accuracy of 
approximately $93\%$, correctly classifying the majority of events, with residual misclassifications visible in the off-diagonal elements of the confusion matrices.

\begin{table}[!htp]\centering
 \scriptsize
 \begin{tabular}{lrrrrcc}\toprule
 Model &Class &Precision &Recall &F1-Score &Accuracy (overall) \\\midrule
 \hline
 R-ConvNet &GRB &0.94 &0.94 &0.94 &{ } \\
 &TGF &0.91 &0.99 &0.95&0.93 \\
 &SGR &0.94 &0.90 &0.92&{ } \\
 &SFLARE &0.95 &0.88 &0.91&{ } \\
 \hline
 MI-DCL &GRB & 0.94 &0.91 & 0.92 &{ }\\  
 &TGF &0.93 &0.98 & 0.95& 0.93 \\
 &SGR & 0.95  & 0.91&0.93 &{ }\\
 &SFLARE &0.91   & 0.92& 0.92&{ }\\
 \bottomrule
 \end{tabular}
 \caption{The classification report for both the R-ConvNet and MI-DCL models on transient classification are give above. The table includes the precision, recall and F1-score for each class of transient: GRB, TGF, SGR, and SFLARE along with the overall Accuracy of each model. Both models show consistently high scores and balanced precision and recall.}
 \label{tab:class_report}
 \end{table}

To account for instances in the test data that do not closely resemble any of the trained classes, we introduced a fifth class, labelled 
“UNCERTAIN”. 
This allows the model to flag events that differ significantly from the known classes, based on their classification confidence. Specifically, if the softmax confidence for all classes
falls below $60\%$ (refer Appendix \ref{appendix:threshold} for more details), we categorise that instance under the UNCERTAIN class, effectively treating it as dissimilar to any of the trained categories. This strategy enables the model to handle out-of-
distribution or ambiguous events, providing a practical mechanism for identifying potentially novel or misclassified transients. To illustrate the impact of this approach, we present two sets of confusion matrices: 
the standard four-class classification without the uncertainty filter, and the expanded five-class classification after applying the confidence 
threshold (Figure \ref{fig:conf_mat}). This comparison highlights how the model performs both with and without the incorporation of the uncertainty criterion. Both our models classified around 2.56$\%$(MI-DCL) and 1.96$\%$(R-ConvNet) of the test data into the uncertain class.  

Continuing from the above results, we further examined the classification performance by visualising the learned feature representations using two-dimensional UMAP (Uniform Manifold Approximation and Projection; \citealt{McInnes2018}) projections for both models. Figure \ref{fig:umap} shows the UMAP plots for the MI-DCL model (panel a) 
and the R-ConvNet model (panel b). These visualisations provide an intuitive understanding of how well the models are able to distinguish between different transient classes in the high-dimensional feature space. 
Events correctly classified by the models are represented by distinct colours corresponding to their true class labels (GRB, TGF, SGR, and SFLARE), while events classified as 'UNCERTAIN' are shown in brown triangles. 

There are instances of 'mislabeled' where the predicted class assigned by the model does not match the true class label. In the UMAP 
projection shown in Figure 8, such events are marked in purple to highlight disagreements between prediction and true class, while distinct marker 
shapes indicate false negatives (FN) for the GRB, SFLARE, SGR and TGF classes. In general, the position of a point in the UMAP space reflects similarity in the 
learned feature representation and therefore indicates the class toward which the model tends to 
assign that event. Thus, markers appearing within a given cluster correspond to events whose features are most similar to that class, irrespective of their 
true labels. The purple points represent false positives for the cluster class and, at the same time, false negatives for their respective true 
classes. For instance, within the SGR cluster, the marked events correspond to false negatives of GRB and SFLARE, indicating true GRB and SFLARE events 
whose representations lie closer to SGR and are consequently predicted as SGR. The presence of these points highlights overlap in the learned feature 
space and indicates regions where class separation is not fully distinct. The fraction of test data that were misclassified can be determined from the off-
diagonal elements of the confusion matrices presented in Figure~\ref{fig:conf_mat}.



The clustering observed in the UMAP plots indicates that both models effectively capture the distinguishing features of each transient 
class. Notably, the UNCERTAIN events predominantly appear near the boundaries of these clusters, reflecting regions where the model finds 
it challenging to make confident predictions. This further validates the utility of the uncertainty class in identifying ambiguous or out-of-distribution samples.

\section{Discussion}

\subsection{Analysis of Misclassified Events and Model Limitations}

\begin{figure*}[ht]
    \centering
    \subfloat[]{\includegraphics[width=0.45\textwidth]{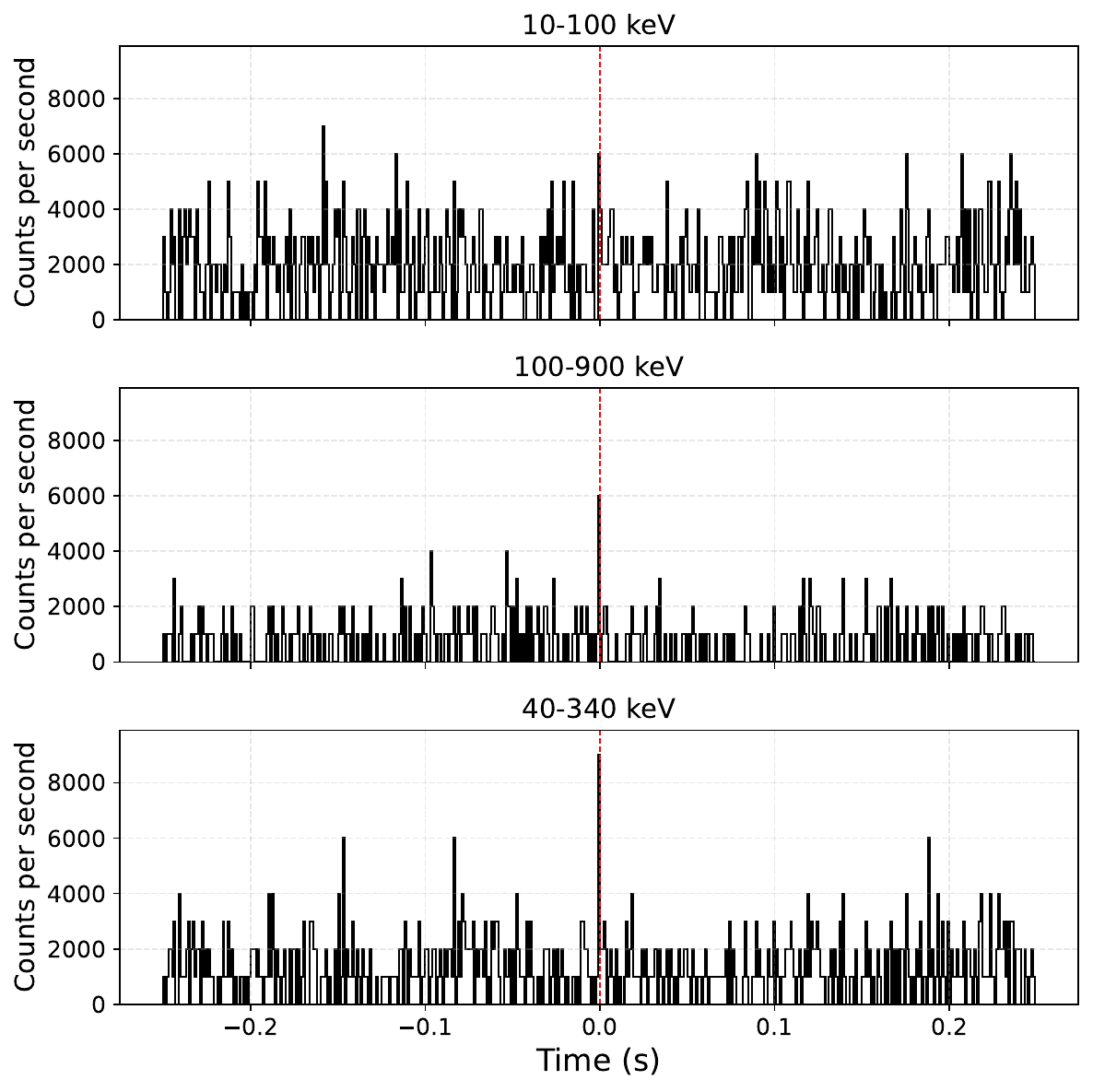}}\hfill
    \subfloat[]{\includegraphics[width=0.45\textwidth]{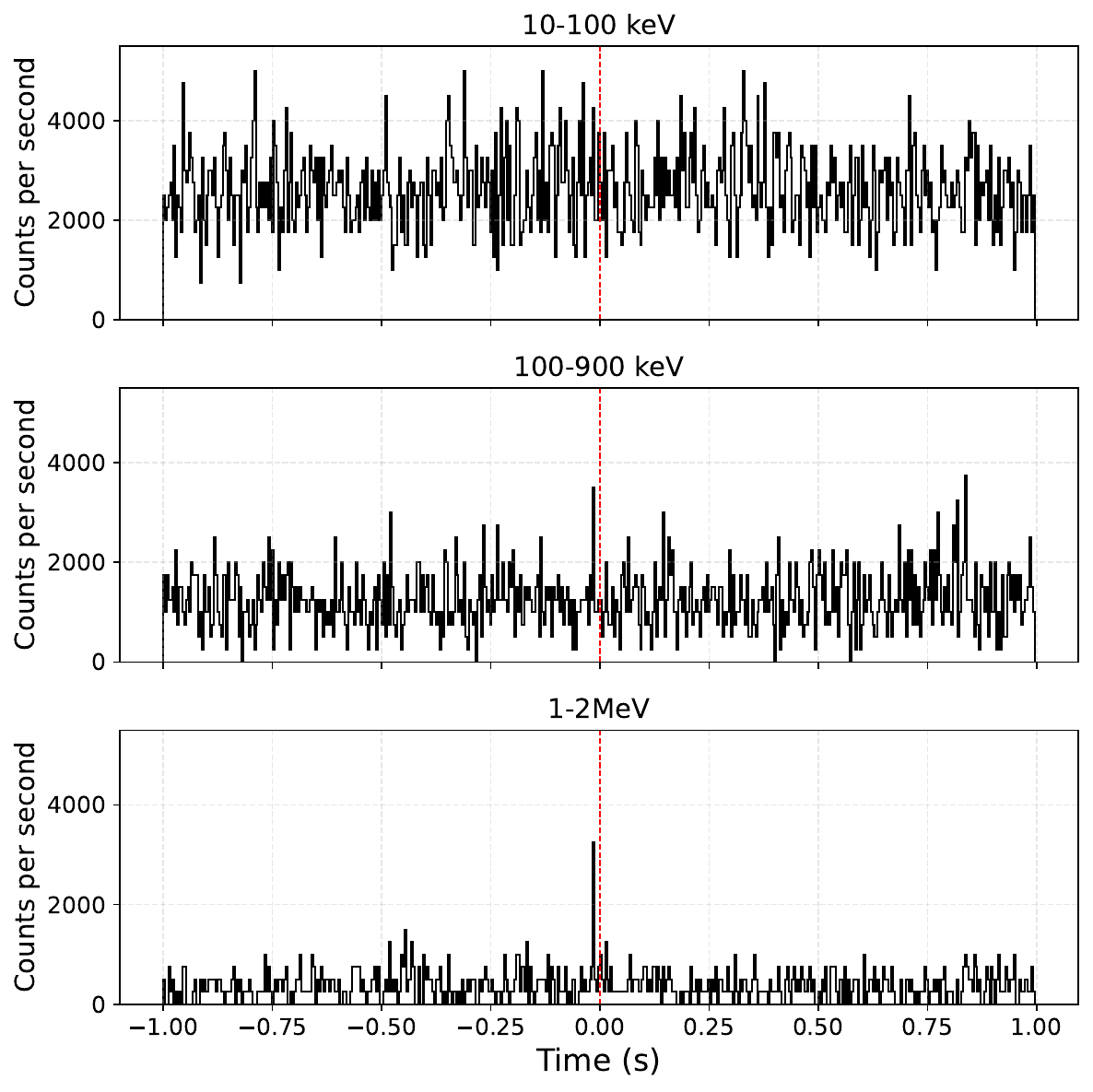}}\hfill
    \subfloat[]{\includegraphics[width=0.55\textwidth]{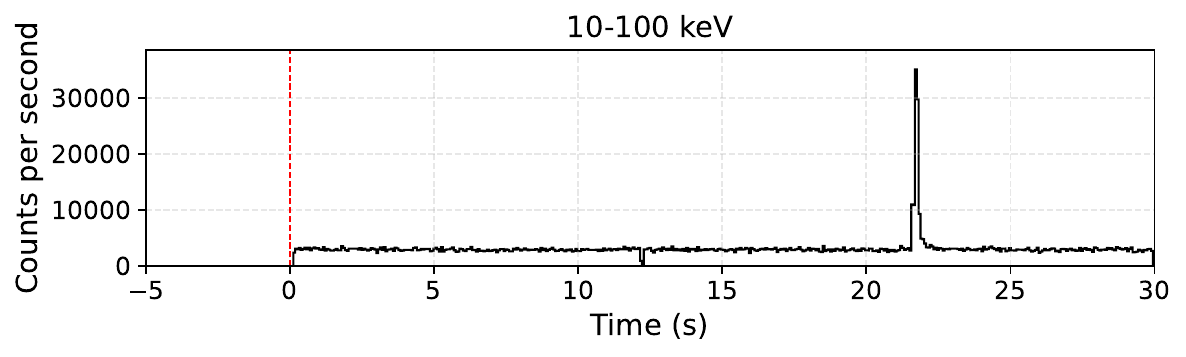}}
    \caption{Our deep learning models failed to identify the above type of events. (a) The above TGF's signal is somewhat visible only in the energy band 40 - 340 keV (shown above) not in the energy ranges chosen for study; (b) This event is visible only in the 1-2 MeV (shown above) while our study is restricted to NaI whose energy range is till 900 keV only;  
    (c) This GRB has its peak signal occurring nearly 20s after the trigger time.}
    \label{fig:zerosig}
\end{figure*}

Despite the overall strong performance of our models, achieving over $93\%$ accuracy, some instances of misclassification were consistently 
observed. A detailed inspection of these misclassified cases revealed several contributing factors. One of the main causes of uncertain classifications appears to be weak signal strength 
, where the transient does not produce a 
sufficiently strong count rate within the selected energy ranges, making it difficult for the model to distinguish it from other classes (Figure 
\ref{fig:zerosig}a). 
Additionally, some events with dominant emission in the MeV range become unclassifiable due to our chosen NaI detectors which is restricted till 900 keV (Figure 
\ref{fig:zerosig}b). Uncertainties can also arise from inconsistent backgrounds or signals not centered on the trigger (Figure 
\ref{fig:zerosig}c). In such cases, the triggering signal may represent a weak precursor, while the main high-count event occurs several seconds later, beyond the time window selected by our current algorithm.


Another source of misclassification was the occurrence of atypical light curve patterns that did not conform to the typical temporal morphology 
seen in their respective classes. 
For example, certain GRBs exhibited extended emission (higher T90 durations) that overlapped with typical SFLARE durations, leading to occasional misclassification between these two classes. 
Similarly, some SGR events with shorter, sharper profiles were misidentified as TGFs, owing to their overlapping short-duration features and comparable count structures in the binned light curves.



Our analysis also suggests that class imbalance, particularly the 
relatively smaller number of SGR events in the training data, might have contributed to reduced classification robustness for this class. 
Increasing the volume and diversity of training examples for underrepresented classes, especially SGRs, would likely improve classification fidelity.

Lastly, we note that our current model architecture is based solely on time-series photon counts in two energy bands. The addition of 
contextual features, such as localisation information (Right Ascension and Declination), could potentially help resolve classification 
ambiguities. This is particularly applicable for classes like SFLARE and TGF, which are confined to specific astrophysical or geophysical 
origins, unlike GRBs which are isotropic. Incorporating such auxiliary 
features in future model iterations may improve the classification accuracy further, particularly in cases where light curve morphology alone is insufficient to distinguish between classes.

\subsection{Comparison with Existing Classification Strategies}
Existing classification methods on {\it Fermi}-GBM rely on threshold-based algorithms, which are efficient but can miss weak or atypical events (see section \ref{Uncertain_disc} for more discussion).
Apart from this there were some efforts to identify the GRB using machine learning and neural networks. \cite{Dimple_etal_2024} used 
DBSCAN clustering and Dynamic Time Warping to detect GRB from the AstroSat CZTI data; their module showed the capability to identify both long and short types of GRBs. While the module enhanced the fast 
automated GRB detection, it is used only for GRB detection and not for any other transients. The model is less sensitive to the short GRBs, 
and additionally, there is a trade-off between the template bank size and DTW computation efficiency.

A model proposed by \cite{PengZhang2024} adopts a similar approach to ours. Their CNN-based framework uses count maps from the 12 NaI 
detectors for binary GRB classification, considering all energy channels. They developed three architectures—plain CNN, ResNet, and 
ResNet-CBAM—each trained with time resolutions of 64, 128, and 256 ms 
over a 120 s window. By generating count maps from each triggered NaI detector, they expanded their GRB dataset from 3083 to 6330 samples. The non-GRB set, drawn from 120 s segments of {\it Fermi} GBM background 
data, comprised 10000 randomly selected examples from 40000 candidates. Their training, validation, and test sets contained 
3082/6000, 1507/2000, and 1741/2000 GRB/non-GRB samples, respectively. The best performing model, ResNet-CBAM (64 ms), achieved $96.57\%$ accuracy, which increased to $96.81\%$ after model fusion.


Furthermore, \citealt{Abraham2021} developed an ML-based method for automated detection of GRB candidates in the 60–250 keV range using 
AstroSat CZTI data. Their approach applied density-based spatial clustering to identify excess power, followed by unsupervised 
hierarchical clustering to classify distinct light-curve patterns.

Thus, several deep learning approaches have been developed to classify events within a specific transient type, such as GRBs, or to 
distinguish particular trigger types from others. However, a framework that classifies multiple kinds of high-energy transients directly from 
time-series data while also identifying ambiguous or potentially novel events remains largely unexplored. The current work, thus, addresses to bridge this gap.

\begin{figure*}[ht]
    \centering
    \subfloat[]{\includegraphics[width=0.45\textwidth]{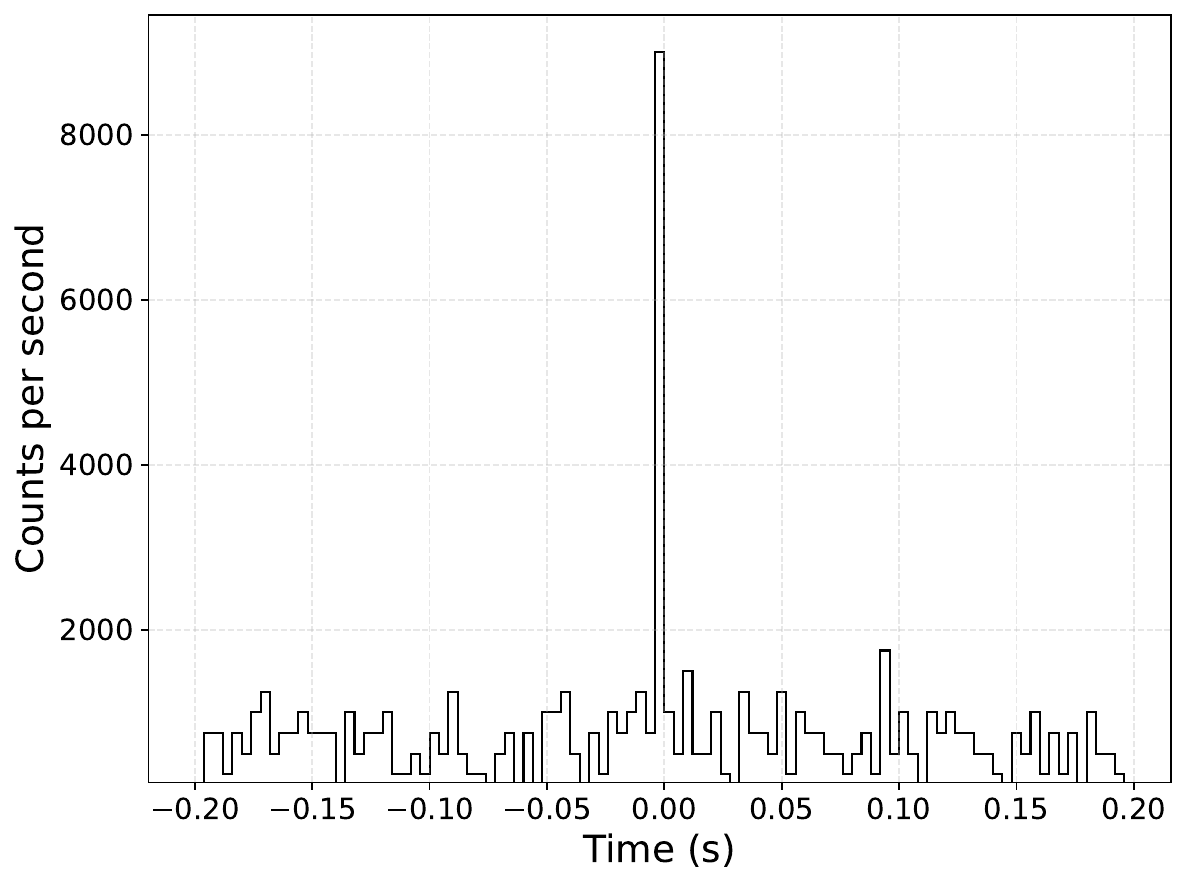}}\hfill
    \subfloat[]{\includegraphics[width=0.45\textwidth]{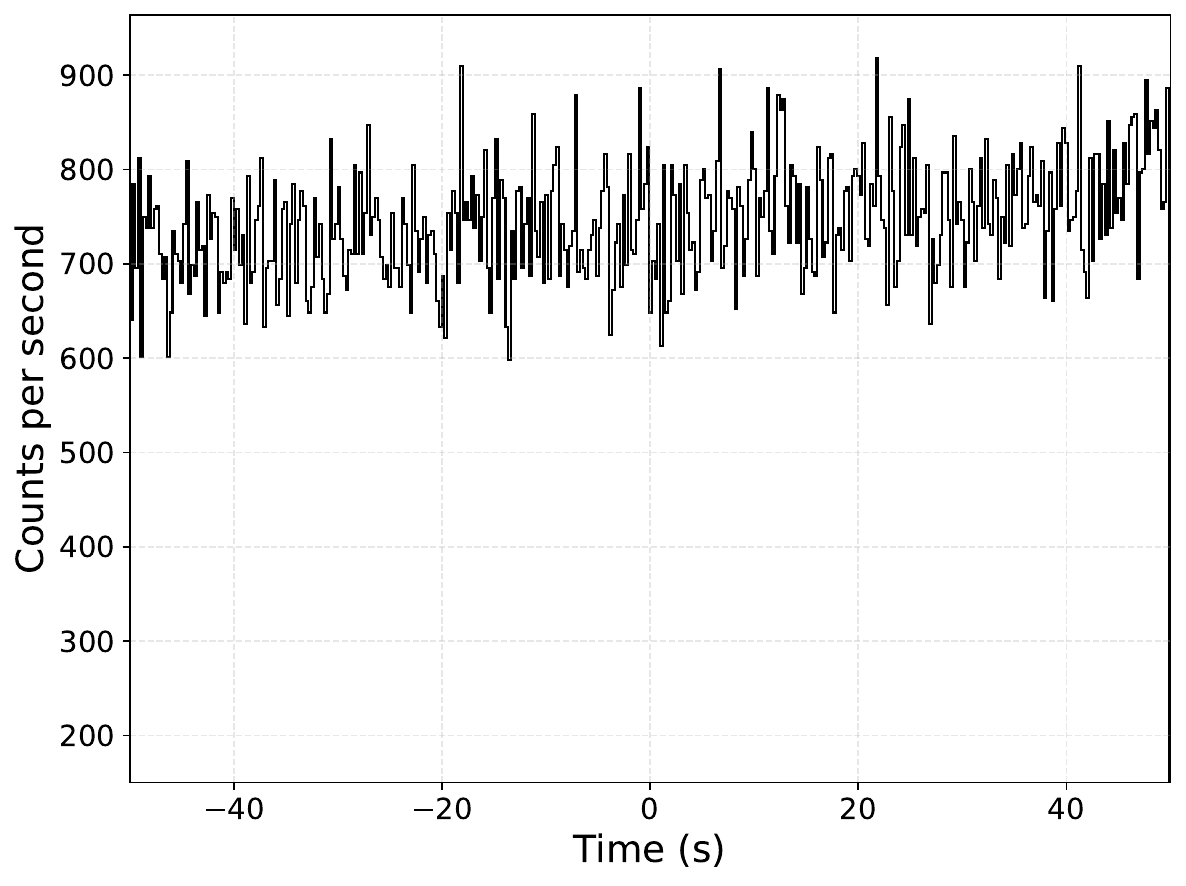}}
    \caption{The lightcurves of a few triggers which were unclassified by {\it Fermi} GBM are shown above. (a) UNCERT bn191025926, however, got classified as TGF by both R-ConvNet and MI-DCL (100-900keV plot) models. (b) UNCERT bn220108026 was unclassified by R-ConvNet (100-900keV plot). No signal was clear in any time and energy binning.}
    \label{fig:Fermiuncerts}
\end{figure*}


\subsection{Computational Efficiency and Resource Utilisation}
\label{sec:eff}

Our models take a very short time for training, within 1 minute using the P100 GPUs on Kaggle\footnote{\url{https://www.kaggle.com/}}. 
And for any new transient data instance, it can predict the class under 0.05 seconds. We have used the free GPU (GPU P100) provided by Kaggle Notebooks\footnote{\url{https://www.kaggle.com/docs/notebooks/}} to train our models, which has reduced our model training time significantly. 
The GBM Flight Software provides an initial onboard classification within approximately 1–5 seconds \citep{Meegan_2009}. However, subsequent ground-based confirmation, which involves 
which involves additional analysis (without detailed spectral study) and human oversight, typically requires approximately 0.5-1 hour. In contrast, once the data are available on the ground, our 
neural network models including data processing can generate classifications in less than 1.6 seconds.  The data pre-processing in Python takes nearly 1.5 
seconds (CPU - AMD Ryzen 5 3550H), but this overhead could be significantly reduced through optimization or by implementing the pipeline in a faster language such as C/C++. 

Occasionally, the data processing fails to identify a signal. This can be because the signal is not clearly visible in the chosen energy ranges (10-100 keV and 100-900 keV) or the signal is displaced from the trigger time. These abnormalities could be accounted for using a more complex algorithm in future works. For now these do not constitute a significant part of the data set and are hence neglected.
    

\subsection{Interpretation of Uncertainty
Classes Reported by {\it Fermi} and our models}
\label{Uncertain_disc}
In the {\it Fermi} GBM trigger catalogue, approximately $6\%$ of events remain unclassified, having failed to be confidently categorized by the onboard GBM flight software as well as by subsequent ground-based analyses involving human review.
To assess the performance of our models on this ambiguous dataset, we applied our models to 50 of these unclassified events. MI-DCL provided a confident ($>0.6$) prediction for all these events, while R-ConvNet failed to classify 7 of these events and interestingly, a majority of the data, around $60\%$ is getting classified as TGF by both models. 
The lightcurves of some of these triggers are shown in Figure \ref{fig:Fermiuncerts}. This shows the potential for these neural network models to help classify this data when the {\it Fermi} GBM's methods fail. Even if the prediction is not always right it provides a good starting point for future investigation of the trigger. 



In our classification model, the fifth category labeled  'UNCERTAIN' 
serves as a safeguard against forcing low 
confidence predictions (softmax $< 0.6$) into one of the four known transient classes, thus providing a quantitative measure of classification reliability. The incorporation of the UNCERTAIN class proved insightful. 
By applying this thresholding mechanism, we were able to segregate ambiguous or novel events that did not exhibit sufficient similarity to any of the four 
trained classes. This approach reduces false positives in the primary classes and offers a systematic method to flag anomalous or outlier events.

A closer examination of events classified under the UNCERTAIN category revealed that a subset of these cases exhibited extremely low signal 
strengths, while others showed unusual or atypical temporal profiles, such as a highly varying background or an uncentred peak with respect to the trigger time (Figure 
\ref{fig:zerosig}c). Additionally, some events may correspond to transient phenomena not represented within our training set. This underscores the utility of the 
UNCERTAIN class as an effective tool for flagging outliers, highlighting anomalous behaviour, and potentially uncovering either rare or entirely new types of high-energy transients, as well as exceptional cases within 
known transient classes. Focused follow-up analyses of these UNCERTAIN events could offer valuable insights into the broader diversity and underlying complexity of transient populations observed by {\it Fermi} GBM.


Furthermore, we note that in both the standard {\it Fermi} classification framework and our models, events that remain unclassified are typically those with lower signal-to-noise ratios or atypical temporal behaviour within 
the studied energy ranges. However, while the FERMIGTRIG catalogue leaves approximately $6\%$ of triggers unclassified, our classifiers show a substantially improved 
performance. For the subset of 713 events evaluated in this study, comprising 663 events from our test set and 50 triggers unclassified by {\it Fermi} GBM, our models leave only 
$2.81\%$ (R-ConvNet) and $2.38\%$ (MI-DCL) of events unclassified. This significant reduction underscores the enhanced 
sensitivity and robustness of our classifiers relative to the existing {\it Fermi}-GBM classification methodology.

To further assess performance on challenging cases, we evaluated our models on GRB events particularly a sample of $51$ GRBs that were reported by {\it Fermi} with reliability scores less than $0.6$. On this low-confidence 
cases, R-ConvNet correctly classified $50\%$ of events with an average confidence of $0.88$, while MI-DCL achieved $78\%$ accuracy with a mean confidence of $0.89$. These 
results indicate that our models retain strong discriminative capability even for near-threshold events. A more comprehensive evaluation incorporating a larger sample of 
human-classified edge cases, such as low signal-to-noise events or transients with ambiguous temporal characteristics, would allow for a more rigorous assessment of model 
performance in this regime. Including such events in the training set is expected to enhance model stability for near-threshold cases and improve generalization to new data.


\section{Summary}
To summarize, we have developed and implemented two R-CNN based deep learning models for the supervised classification of four major 
types of high energy transients: GRBs, TGFs, SGRs, and SFLAREs, using time series light curve data from the Fermi Gamma-ray Space Telescope.

Both models achieved an overall accuracy of approximately $93\%$, with minimal misclassification. A fifth `uncertain' category was introduced for events with softmax confidence below $60\%$ across 
all classes, accounting for about $1–2\%$ of the test data. UMAP visualizations reveal that these uncertain events predominantly lie near the cluster boundaries of different trigger types, reflecting their inherent ambiguity.

This work thus presents two complementary deep learning frameworks capable of classifying diverse high energy transients directly from 
time series data while identifying ambiguous or potentially novel events. With an efficient processing time of only $\sim 1.55$ 
seconds per event, these models has the potential to be integrated into future observation pipelines to enable rapid classification and facilitate timely, coordinated multi-wavelength and multi-messenger follow-ups of high energy transients.

\ack{
This research has made use of {\it Fermi} data obtained through High Energy Astrophysics Science 
Archive Research Center Online Service, provided by the NASA/Goddard Space Flight Center. This work utilized various software such as 3ML \citep{threeML_}, NAIMA \citep{naima}, PYTHON 
(\citealt{Python_1}), ASTROPY (\citealt{Astropy_2013,Astropy_2018,Astropy_2022}), NUMPY (\citealt{Numpy_1}), SCIPY (\citealt{Scipy_1}), MATPLOTLIB (\citealt{Matplotlib_1}), FTOOLS 
(\citealt{ftools_1}), Tensorflow (\citealt{tensorflow2015-whitepaper}) etc. This research utilized computational resources made 
available through Kaggle’s cloud infrastructure.}

\funding{
S.I. is supported by DST INSPIRE Faculty Scheme (IFA19-PH245), SERB SRG Grant (SRG/2022/000211) and ANRF ARG Grant (ANRF/ARG/2025/000380/PS).}

\roles{
S.I conceived the study, designed the methodology, and supervised the overall research. The model development 
and data processing were carried out by A.A.J and K.G.S contributed to data analysis and visualization. A.A.J, K.G.S and S.I interpreted the results and 
prepared the manuscript. S.B. provided expert guidance on machine learning and deep learning methods, contributed to model evaluation, and assisted in 
refining the analysis and interpretation of results. All authors reviewed and approved the final version of the paper.}

\data{The data analysed in this work are publicly available from the {\it Fermi} GBM Trigger Catalog (\url{https://heasarc.gsfc.nasa.gov/w3browse/fermi/fermigtrig.html}). The data processing scripts and trained models developed for this study are available at: 
\url{https://github.com/AstroBurst/Multivariate_Time_Series_Classification_of_Fermi-Detected_Gamma-Ray_Transients}}

\appendix

\section{Appendix : Data Processing}
\label{appendix:data_processing}

 As discussed in sections 2.1 and 2.2, the photon counts from the 3 brightest detectors are merged to produce light curves in 2 energy channels and 7 bin sizes.
 The photon arrival times\footnote{The arrival times of the photons and the trigger time of the burst are reported in Mission Elapsed Time (MET) which for {\it Fermi} is the number of seconds since 00:00:00 UTC on January 1, 2001 (UTC).} 
were first adjusted by subtracting the trigger time, effectively setting the trigger time to zero.\\

For each of the seven bin sizes, the bin edges were calculated using the following equations:
\begin{align*}
    \text{Starting time} &= \text{trigger time} - (dno \times r \times i), \\
    \text{Stopping time} &= \text{trigger time} + (dno \times (1 - r) \times i),
\end{align*}
where $dno = 500$ is the total number of bin edges, $r$ is the ratio of pre-trigger to post-trigger bins, and $i$ is the bin size. When the counts in the TTE files did not fully span the defined time range, some bins were only partially populated. To ensure consistency, such bins were excluded, and the light curves were padded by replacing the non-zero counts at the extremes with zeros, effectively limiting each light curve to 500 bins centered around the trigger time.\\

Following this, the Bayesian block algorithm \citep{scargle2013studies} with a false alarm probability ($p_0 = 0.01$) was applied to each light curve allowing to identify the start and stop times of each transient, as well as facilitated the identification of pre- and post-transient background regions in each light curve. The background was then modeled and subtracted from the signal. \replaced[]{}{The schematic diagram depicting the data processing is shown in Figure \ref{fig:data_processing}}. 
If for any reason the algorithm failed to identify a significant signal in any particular energy or time binning, the data from that binning was omitted to prevent the model from learning from noise.

The background was modeled using a polynomial fit (up to the fourth order was investigated) in the identified pre and post transient time intervals. The best fit was identified using reduced chi-square closest to 1. Subsequently, the polynomial was subtracted from the lightcurve. This procedure was repeated across all seven binnings and both energy ranges generating 14 light background-subtracted light curves.  All data processing was performed using custom scripts developed in Python\footnote{\url{https://github.com/AstroBurst/Multivariate_Time_Series_Classification_of_Fermi-Detected_Gamma-Ray_Transients}}.



\section{Appendix : Threshold Selection}
\label{appendix:threshold}
\begin{figure}[htbp]
\centering

\subfloat[(a) R-ConvNet confidence distribution]{
\includegraphics[width=\linewidth]{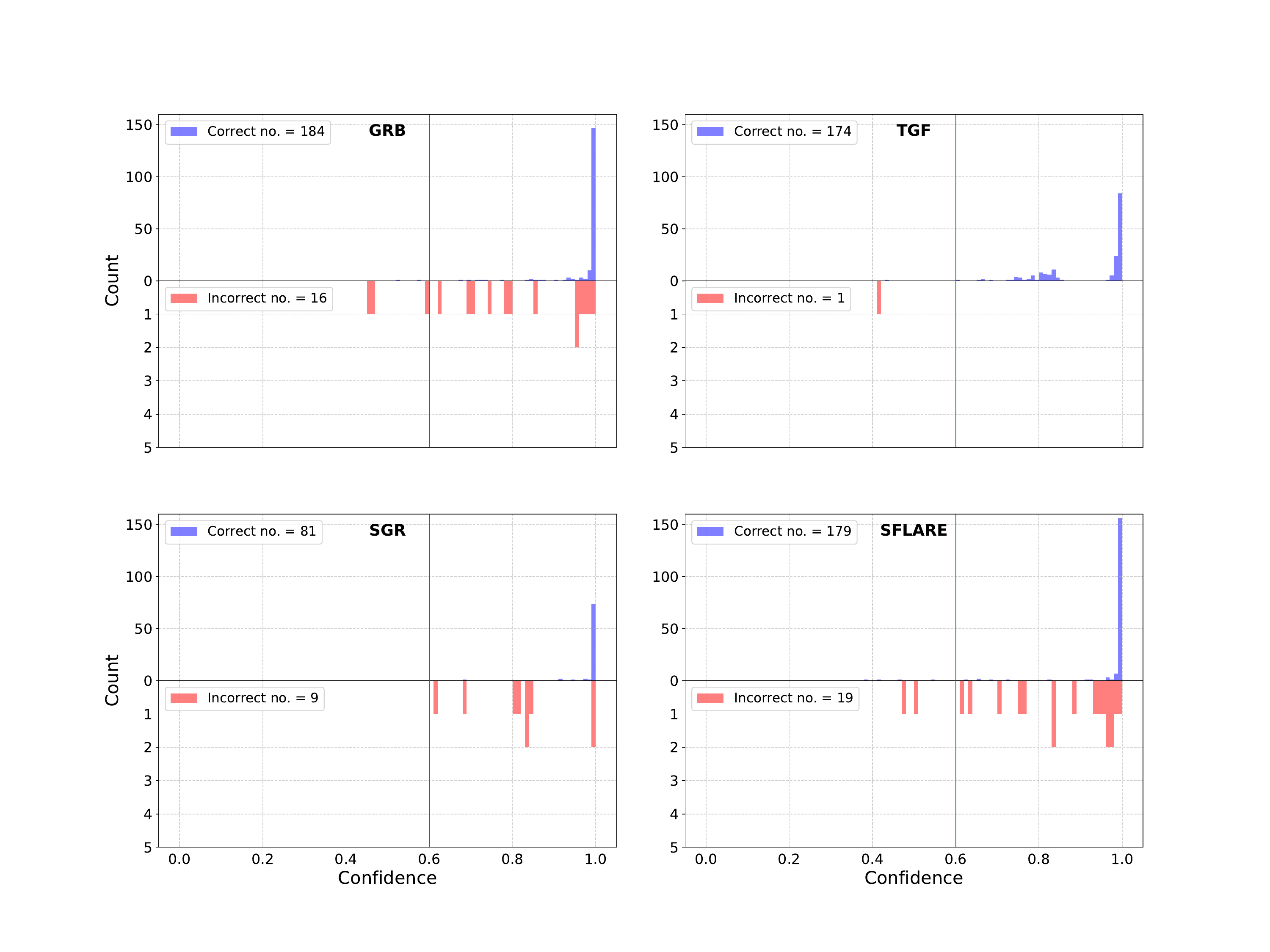}
\label{fig:distributions1}
}

\vspace{-0.5cm}

\subfloat[(b) MI-DCL confidence distribution]{
\includegraphics[width=\linewidth]{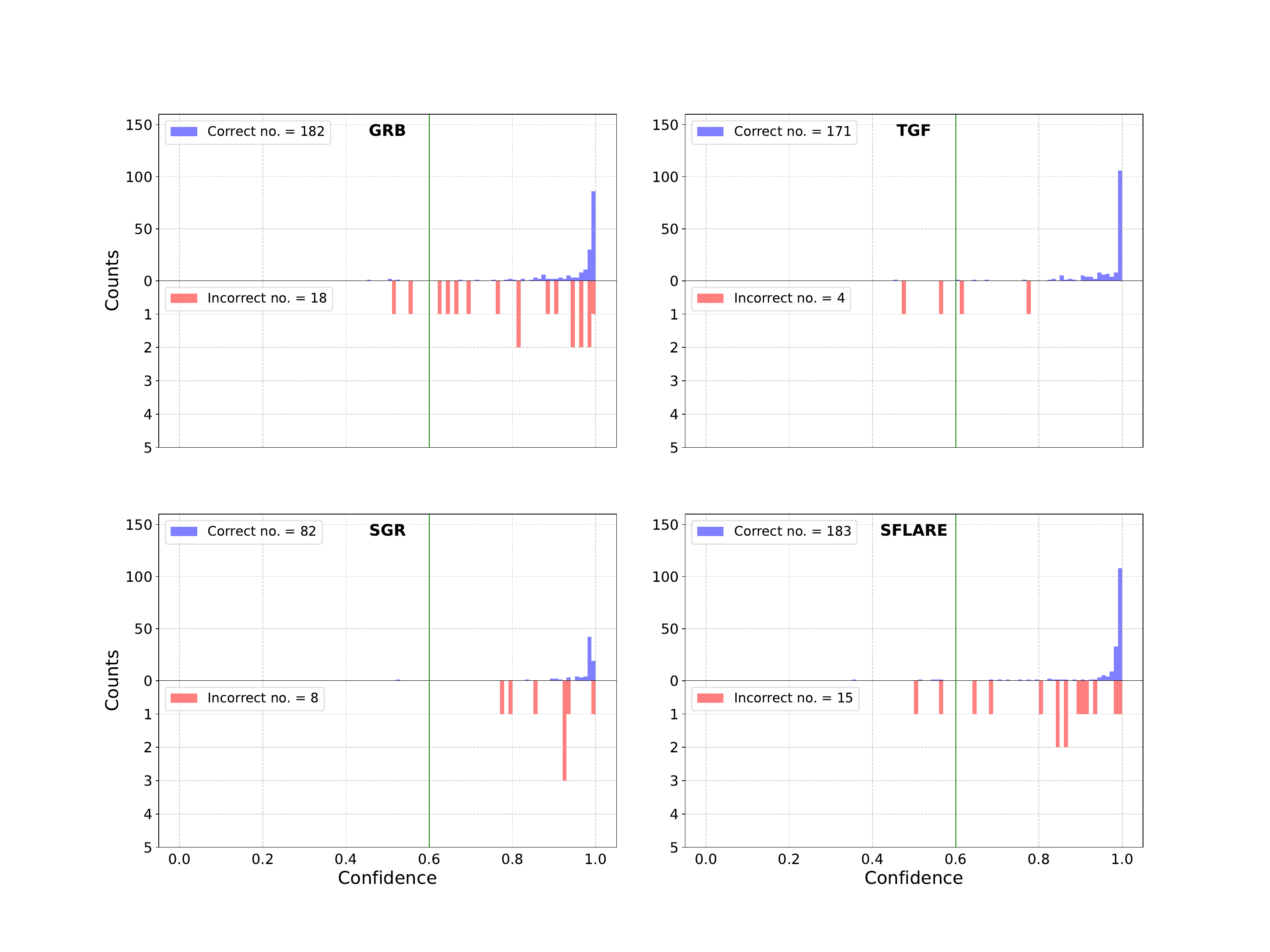}
\label{fig:distributions2}
}

\caption{The confidence distributions for all four classes obtained from the R-ConvNet and MI-DCL models are shown in the upper (a) and lower (b) panels, respectively. In each plot, the upper histogram (blue) represents the confidence values for correctly classified data, while the lower histogram (red) corresponds to misclassified data.}
\label{fig:distributions_combined}

\end{figure}

We selected a threshold value of $60\%$ based on the distribution of classification confidences (i.e softmax confidence obtained for each classification) shown in Figure \ref{fig:distributions_combined}. Most data points were 
classified with high confidence, typically in the $80–90\%$ range. However, setting a higher threshold (e.g. $70–80\%$) would risk labeling genuinely confident predictions as uncertain. Conversely, a lower threshold (e.g. $40–50\%$) 
could result in incorrectly treating uncertain or confused predictions as confident classifications. Therefore, adopting 
a conservative threshold of $60\%$ provides a reasonable balance, minimizing the mislabeling of both uncertain and confident instances.





\end{document}